%% file: draft.tex
\documentclass[aps,prd,twocolumn,showpacs,superscriptaddress,amsmath,longbibliography, nofootinbib]{revtex4-1}

\usepackage{graphicx}
\usepackage{dcolumn}
\usepackage{bm}
\usepackage{epstopdf}
\usepackage{xspace}

\usepackage{hyperref} % add hypertext capabilities 
\usepackage[usenames]{color} % color for proofread
\usepackage{ulem}
\usepackage{multirow}
\usepackage{enumitem}
\usepackage{enumerate}

\RequirePackage{lineno}

\graphicspath{{figures/}}

\begin{document}

%\pagewiselinenumbers
%\linenumbers

\title{
%{\Large NOT FOR DISTRIBUTION -- ONLY FOR INTERNAL USE}\\
%\vspace{0.2cm}
Search for nucleon decay into charged antilepton plus meson in 0.316 megaton$\cdot$years exposure of the Super-Kamiokande water Cherenkov detector}

\input{authors-20170315}

\date{\today}

\begin{abstract}

We have searched for proton decays into a charged antilepton ($e^+$, $\mu^+$) plus a meson ($\eta$, $\rho^0$, $\omega$) and neutron decays into a charged antilepton ($e^+$, $\mu^+$) plus a meson ($\pi^-$, $\rho^-$) using Super-Kamiokande (SK) I-IV data, corresponding to 0.316 megaton$\cdot$years of exposure. 
%Several analysis improvements have been implemented and 2.26 times data has been used with respect to the previously published results. 
This measurement updates the previous published result by using 2.26 times more data and improved analysis methods.
No significant evidence for nucleon decay has been observed and lower limits on the partial lifetime of the nucleon are obtained. The limits range from 3$\times$10$^{31}$ to 1$\times$10$^{34}$ years at 90\% confidence level, depending on the decay mode.

\end{abstract}
%\pacs{12.10.Dm,13.30.-a,12.60.Jv,11.30.Fs,29.40.Ka}

\maketitle
%\tableofcontents

%\newpage

\section{Introduction}

\subsection{Grand Unification Theories}

The Standard Model of particle physics, based on the
$\mathrm{SU}(3)_{c}\times\mathrm{SU}(2)_L\times\mathrm{U}(1)_Y$ gauge
symmetry, seems to be an incomplete description of Nature's structure
of matter and forces. The standard model fails to include massive
neutrinos or provide a satisfying explanation for charge quantization
or the convergence of gauge couplings at high energy scales. These
issues are addressed in Grand Unified Theories (GUTs), which embed the
standard model in a single unifying gauge group such as
$\mathrm{SU}(5)$ \cite{Georgi:1974sy} or $\mathrm{SO}(10)$
\cite{Fritzsch:1974nn}. While numerous specific realizations of GUTs
exist (see \cite{Nath:2006ut} for review), as a universal feature they
predict nucleon decay, and therefore, baryon number violation. The
gauge coupling unification scale is around $10^{15-16}$ GeV,
which is unreachable with accelerators but can be explored by virtual
processes that induce nucleon decay with lifetimes of $10^{32-35}$
years. The lower end of this lifetime range may yield observable event
rates in long exposures of a large water Cherenkov detector such as
Super-Kamiokande.

The two benchmark nucleon decay modes $p \rightarrow e^+\pi^0$ and
$p\rightarrow\overline{\nu}K^+$ are favored channels
\cite{Senjanovic:2009kr} in non-supersymmetric GUTs and TeV-scale
supersymmetric GUTs, respectively. However, depending on the model,
the branching ratios for nucleon decay channels vary and different
decay modes can dominate or be comparable to the favored channels. For
example, in $\mathrm{SU}(5)$ the rate for $n \rightarrow e^+\pi^-$ is
twice that of $p \rightarrow e^+\pi^0$~\cite{Langacker:1980js};
however the latter mode has a stronger experimental signature. It is
conceivable that rates for channels with heavy non-strange mesons
($\eta$, $\rho$, $\omega$) are comparable in magnitude to those with
pions~\cite{Machacek:1979tx,*Gavela:1980at,*Donoghue:1979pr,*Buccella:1989za}.
In the lifetime limit where the expectation for any single channel is
one event or less, searching in multiple decay modes enhances the
chance that one observes any nucleon decay at all. Finally, observing
proton decay from a non-standard channel in the absence of a signal
from $p\rightarrow e^+ \pi^0$ might hint at some exotic scenario.

In this work, we describe nucleon decay searches for which the final state
is comprised of two particles: a charged antilepton, $e^+$ or $\mu^+$,
and a non-strange meson. These nucleon decays violate baryon number,
$B$, but conserve baryon number minus lepton number, $B-L$. This paper
in combination with our recent publication searching for $p \rightarrow
e^+\pi^0$ and $p \rightarrow \mu^+\pi^0$~\cite{Miura:2016lpi0},
has covered all such two-body decay modes.

%For example, in the model of \cite{Carone:1995xw}, based on
%non-trivial flavor symmetry and supersymmetry and without Grand
%Unification, the dominant channel is $p\rightarrow e^+K^0$, in
%contrast to the usual SUSY GUT predictions.

\subsection{The Super-Kamiokande Experiment}

Super-Kamiokande (SK) is a large water Cherenkov detector located in
the Kamioka mine under about 1~km rock overburden (2.7 km water
equivalent) at the Ikenoyama mountain in
Japan~\cite{Fukuda:2002uc,*Abe:2013gga}.  It is composed of a 39.3~m
diameter by 41.4~m high vertical cylinder lined with 50-cm
photomultiplier tubes (PMTs) facing the inner volume. The SK
experiment has been collecting data since 1996 in four different
experimental phases denoted as SK-I to SK-IV. The inner detector photo
coverage is around 40\% in SK-I, SK-III, and SK-IV but was 19\% in
SK-II. New front-end electronics called
QBEE~\cite{Nishino:2009zu,*Yamada:2010zzc} was implemented in SK-IV,
which enabled deadtime-free data acquisition of a wide time window of
PMT hit information. This enabled neutron capture to be included
in SK-IV analyses. The SK detector has been extensively tuned and
calibrated~\cite{Fukuda:2002uc,*Abe:2013gga}.

Due to its large fiducial volume (22.5~kilotons, corresponding to
about 7.5$\times$10$^{33}$ protons and 6.0$\times$10$^{33}$ neutrons),
excellent event reconstruction performance~\cite{MS:nim} as well as
long stable detector operation, SK has sensitivity to nucleon decay
generally exceeding past experiments by an order of magnitude or
more. Decays into a charged antilepton and a meson are particularly
well suited for water Cherenkov detector searches, since both the
lepton and the meson or their decay particles are above the Cherenkov
threshold and are visible in most cases.

The search for nucleon decay into a charged antilepton plus a meson
using data from SK-I and SK-II (0.141 megaton$\cdot$years) has been
published previously~\cite{Nishino:2012bnw}. Among those nucleon decay
modes, the $p \rightarrow (e^+$, $\mu^+) \pi^0$ searches have been
updated recently with additional data from SK-III and SK-IV along with
several analysis improvements~\cite{Miura:2016lpi0}. No evidence has
been observed for any nucleon decays and the corresponding lifetime
limits were set for these channels.

In this work, we update the searches for proton decay into a charged
antilepton ($e^+$, $\mu^+$) plus a meson ($\eta$, $\rho^0$,
$\omega$) and neutron decay into a charged antilepton ($e^+$,
$\mu^+$) plus a meson ($\pi^-$, $\rho^-$) using the latest data set
from SK-I to SK-IV, an exposure of 0.316 megaton$\cdot$years.

\subsection{Analysis improvements}

This paper incorporates several major improvements in the analysis
over the previously published results~\cite{Nishino:2012bnw}. These
improvements have already been implemented for $p \rightarrow e^+
\pi^0$ and $p \rightarrow \mu^+ \pi^0$ and recently
published~\cite{Miura:2016lpi0}.

%% The nucleon decay lifetime sensitivity is directly proportional to the
%% detector's exposure for the background free case, but goes as the
%% square root of exposure for the non-zero background case. Hence, it is
%% important to increase the signal efficiency and the background
%% rejection.

%The major analysis improvements in this work, compared to the
%previous study of Ref.~\cite{Nishino:2012bnw}, are the same as those
%implemented in the recent $p \rightarrow l^+ \pi^0$ search ($l^+$ =
%$e^+$, $\mu^+$)~\cite{Miura:2016lpi0}.

For some channels, the signal region in the reconstructed total
momentum and invariant mass distribution parameters is separated into
two distinct sub-regions (``boxes"), according to the total
momentum. This permits a statistical separation of bound and free
proton decays for channels that have good invariant mass and momentum
resolution. For this paper, the channels that benefit are the $p
\rightarrow l^+ \eta$, $\eta \rightarrow 2\gamma$ searches 
($l^+$ = $e^+$, $\mu^+$). 
The majority of the free proton decays have total momentum in the lower-momentum box, where the number of expected atmospheric neutrino background
events is negligibly small. The two-box momentum separation is
implemented for these channels in all data sets from SK-I to SK-IV. For
more details see Section~\ref{sec:p2leta}.

Atmospheric neutrino background is often accompanied by neutron
production whereas it is expected that nucleon decay seldom releases a
free neutron~\cite{Ejiri:1993rh}. For SK-IV analysis, a neutron capture
reconstruction algorithm~\cite{Irvine:2014hja} is included which
allows us to reduce the atmospheric neutrino background rate. The
neutron tagging is available, only for SK-IV, thanks to a triggerless
readout scheme that records every hit in a wide time window. Following
a primary event, we search over 435~$\mu$s for coincident hits near
the vertex from the 2.2~MeV gamma ray from neutron capture on
hydrogen, which occurs with a mean capture time of 200 $\mu s$ in
water. Description of the neutron simulation and the tagging algorithm
can be found in Ref.~\cite{Miura:2016lpi0}. In this analysis, the
number of expected SK-IV background events is reduced using neutron
tagging by 30-90\% depending on the nucleon decay modes. A full
summary of the resulting background for each SK period can be found in
Table~\ref{tab:summaryall}.

% ----- Hide.T added (2016/10/26) ------------ The systematic
%uncertainties of the pion interactions in the oxygen nuclear medium
%and the pion interactions in the detector medium have been
%reevaluated~[ref: SK epi0 2016] since the previous analysis of
%Ref.~\cite{Nishino:2012bnw}.

The systematic uncertainties of the pion interactions in the oxygen
nuclei and water have been reevaluated since the previous analysis of
Ref.~\cite{Nishino:2012bnw}. See Section~\ref{sec:syserr} for more
details.

% by comparing NEUT~\cite{{Hayato:2002sd, Mitsuka:2007zz,
% Mitsuka:2008zza} simulation, which is based on the Bertini
% intra-nuclear hadronic cascade model~\cite{Bertini:1970zs}, and the
% pion-nucleus scattering data [ref]. This method provides a more
% reliable evaluation of the systematic uncertainties for pion FSI and
% SI.

% --------------------------------------------

%The new two-box analysis and neutron tagging achieve better discovery
%reach. For example, the 3-$\sigma$ discovery potentials in the proton
%lifetime for $p \rightarrow e^+ \eta$ search are estimated to be
%about 12\% and 20\% higher than the previous single-box analysis
%without neutron tagging, assuming for exposure 0.316
%megaton$\cdot$years (current analysis) and 1 megaton$\cdot$years
%(future), respectively.  The discovery potential is calculated by
%assuming the signal efficiencies, the number of expected backgrounds,
%and their systematic errors for SK-IV as used in this analysis. %The
%same calculation method was adopted in \cite{Miura:2016lpi0}.

The new two-box strategy and neutron tagging provide better discovery
reach. For example, the 3-$\sigma$ discovery potential in the proton
lifetime for $p \rightarrow e^+ \eta$ search is estimated to be
about 12\% higher than the previous single-box analysis without
neutron tagging for the exposure used in this analysis.

\section{Simulation}

\subsection{Nucleon decay}

Nucleon decay Monte Carlo (MC) samples are generated to determine the
search strategy and to estimate the signal efficiencies. The MC
simulation includes: initial nucleon decay kinematics, meson
interactions in the oxygen nucleus for bound nucleon decays, hadron
propagation in water, propagation of particles, and Cherenkov light
emission, absorption, and scattering. The initial nucleon decay
kinematics are identical to those in Ref.~\cite{Nishino:2012bnw},
while updated pion interactions and particle and Cherenkov light
simulations in water are adopted from Ref.~\cite{Miura:2016lpi0}.

\begin{figure}[htbp]
\begin{center}
\includegraphics[width=.9\linewidth]{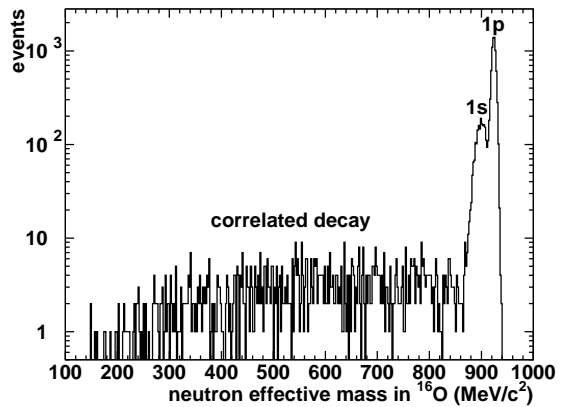}
\end{center}
\caption{ \protect \small The effective mass of the decayed neutron in
  $^{16}$O for $n \rightarrow e^+ \pi^-$ as simulated by Monte Carlo.
}
\label{fig:neffmass}
\end{figure}

In the case of nucleon decay within the oxygen nucleus, the effects
due to Fermi momentum, correlation with other
nucleons~\cite{Yama:corr} (correlated nucleon decay), nuclear binding
energy, and meson-nucleon interactions are all taken into
account~\cite{Nishino:2012bnw}.  The effective mass of the decayed
neutron for $n \rightarrow e^+ \pi^-$ MC is shown in
Figure~\ref{fig:neffmass}. Unlike for the proton decay, there is no
sharp peak at the neutron mass (940~MeV/c$^2$) because all the
neutrons in a H$_2$O molecule are bound in a nucleus and there is no
free neutron decay. The largest and most energetic peak corresponds to
the 1$p$-state; the 1$s$-state is nearby. The nuclear binding energies
are 39.0~MeV for the 1$s$-state and 15.5~MeV for the
1$p$-state~\cite{Nakamura:1976mb}. Correlated neutron decay, which
accounts for the possibility of another nucleon recoiling with the
decaying nucleon, makes the tail in the lower mass region.

The meson-nucleon interactions in an oxygen nucleus are simulated with
the NEUT neutrino interaction
simulation~\cite{Hayato:2002sd,*Mitsuka:2007zz,*Mitsuka:2008zza}.
Mesons generated in an oxygen nucleus may interact with nucleons
before they escape from the nucleus. The position of the nucleon decay
in a nucleus is determined by the Woods-Saxon nuclear density
distribution in the simulation. From this position, $\pi$, $\eta$, and
$\omega$ mesons are tracked within the oxygen nucleus. The lifetime of
the $\rho$ meson is short enough ($\beta\gamma c \tau\sim$0.3~fm) that
it decays immediately inside the nucleus into two $\pi$
mesons. Therefore, the nuclear effects of the $\rho$ meson itself are
not considered in the simulation but those of the secondary pions are
included.
%

% Hide-san's sentences moved to here from syst. error section

\begin{figure}[htbp]
\begin{center}
\includegraphics[width=.9\linewidth]{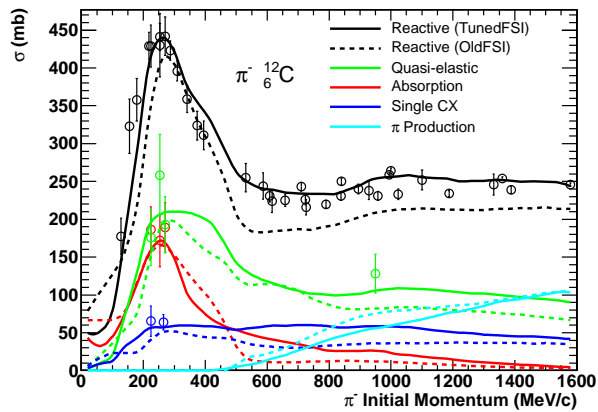}
\end{center}
\caption{ \protect \small
The $\pi^-$ interaction cross section on $^{12}$C as a function of momentum. Open circles indicate existing data~\cite{Giannelli:2000zy,*Saunders:1996ic,*Navon:1983xj,*Bowles:1981my,*Ashery:1981tq,*Ashery:1984ne,*Ashery:1984ne,*Levenson:1983xu,*Jones:1993ps,*Chavanon:1969zu,*Takahashi:1995hs,*Allardyce:1973ce,*Cronin:1957zz,*Aoki:2007zzb,*Rahav:1991vi}, dashed and solid lines show before and after the tune, respectively. The colors correspond to the total cross section and exclusive channels: total (black), quasi-elastic scattering (green), absorption (red), charge exchange ``CX'' (blue), and $\pi$ production (light blue). ``FSI'' stands for final state interaction.
}
\label{fig:pimxsec}
\end{figure}

The pion interaction model in NEUT is a semi-classical cascade model.
The interaction probabilities (mean free paths) for low momentum pions
are calculated with the Delta-hole model by Oset {\it
  et al.}~\cite{Salcedo:1987md}.  The pion interaction probabilities
for high momentum pions are extracted from the $\pi^{\pm}$-nucleus
scattering experimental data.  All the pion interaction probabilities
and cross section parameters have been tuned with various
$\pi^{\pm}$-nucleus scattering data, including C, O, Al,
Fe~\cite{Miura:2016lpi0}, since the previous
paper~\cite{Nishino:2012bnw}.
Figure~\ref{fig:pimxsec} shows the $\pi^-$ cross sections of the external data and MC on $^{12}$C~\cite{Giannelli:2000zy,*Saunders:1996ic,*Navon:1983xj,*Bowles:1981my,*Ashery:1981tq,*Ashery:1984ne,*Ashery:1984ne,*Levenson:1983xu,*Jones:1993ps,*Chavanon:1969zu,*Takahashi:1995hs,*Allardyce:1973ce,*Cronin:1957zz,*Aoki:2007zzb,*Rahav:1991vi}. The tuned MC used in this analysis increased the cross sections at around 500~MeV/c.
%
% reference for 1535?

\begin{figure}[htbp]
\begin{center}
\includegraphics[width=.9\linewidth]{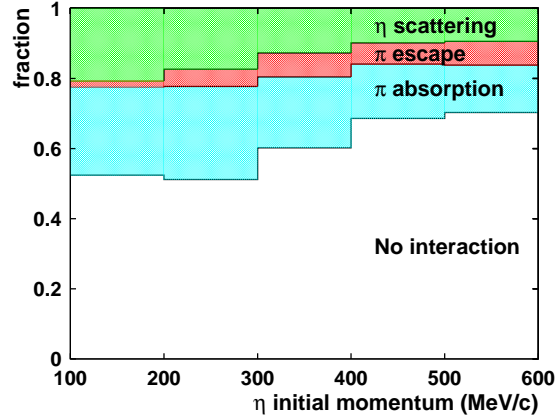}
\end{center}
\caption{ \protect \small
Fraction of meson interactions in the oxygen nucleus in bins of $\eta$ momentum generated by $p \rightarrow e^+ \eta$ MC. 
$\eta$s that exit the nucleus without experiencing any interactions are indicated by the portion labeled ``No interaction''.
If at least one of the pions generated via the baryon resonance of $N$(1535) is absorbed, it is called ``$\pi$ absorption''.
If none of the pions are absorbed, it is called ``$\pi$ escape''.
}
\label{fig:etarainbow}
\end{figure}

The interactions between the $\eta$ and nucleons in the nucleus are
considered through a baryon resonance of $N$(1535) and the
corresponding cross section is calculated by the Breit-Wigner formula.
%The net nuclear effect of the $\eta$ meson in an oxygen nucleus is
%estimated by using this cross section as well as considering the
%nuclear density distribution, nucleon momentum distribution and the
%Pauli exclusion principle effect in an nucleus.
Since pions can be generated by the decay of the resonance, nuclear
effects for them are considered. Figure~\ref{fig:etarainbow} shows the
fractions of the true meson interactions in the oxygen nucleus in the
$p \rightarrow e^+ \eta$ MC.

The $\omega$ meson interactions with a nucleon in an oxygen nucleus are calculated with a boson exchange model~\cite{Lykasov:boson}. 
%The coupling constants and the form factor of this model are fixed by the experimental data.

Uncertainties due to these physics processes are taken into account within the systematic error estimations on the signal efficiencies as discussed in Section~\ref{sec:syserr}.

\subsection{Atmospheric neutrinos}

The background for nucleon decay searches comes from atmospheric
neutrino interactions in water. The standard Super-Kamiokande
atmospheric neutrino simulations~\cite{Honda:2011, Hayato:2002sd,
  Mitsuka:2007zz, Mitsuka:2008zza, Bertini:1970zs, Micap} are used to
estimate the background rate. The model includes an atmospheric
neutrino flux calculation,  neutrino oscillation, and a detailed model
of neutrino-nucleus interaction cross sections. The meson interactions
in the oxygen nucleus are treated similarly for the nucleon decay and the
atmospheric neutrino MC. Simulated data samples equivalent to around
500~years detector exposure are generated for each SK period. The
detector simulation, including Cherenkov light production, scattering,
absorption, and detector response, is identical for the nucleon decay
signal MC and the atmospheric neutrino background MC.

Uncertainties of the neutrino flux, neutrino cross sections,
pion-nuclear interactions as well as hadron propagation in water are
all taken into account in the systematic error estimates on the number
of the expected background events as discussed in
Section~\ref{sec:syserr}.

The neutrino and the pion interactions in the atmospheric neutrino
background for the $p \rightarrow e^+ \pi^0$ search in SK were
experimentally validated using an accelerator neutrino beam and a
1-kiloton water Cherenkov detector~\cite{Mine:2008rt}. The dynamics of
pion production and interactions relevant for the nucleon decay
searches for the SK-type water Cherenkov detector have been verified.

\section{Data reduction and reconstruction}

%\section{Data set and data reduction}

Data from 91.5, 49.1, 31.8, and 143.8 kiloton$\cdot$year exposures, corresponding to the 1489.2, 798.2, 518.1, and 2339.4 live days of the experiment during the SK-I, SK-II, SK-III, and SK-IV phases, are used. Several stages of reduction are applied to suppress the cosmic ray muon background and prepare samples for the physics analyses. The reduction algorithms are identical to those used for atmospheric neutrino analyses and nucleon decay searches~\cite{Ashie:atmosc}.

Fully contained (FC) events are selected by requiring a vertex inside of the 22.5 kiloton fiducial volume (2~m away from the inner detector wall), visible energy greater than 30~MeV, and no hit-PMT clusters in the outer detector. 

%\clearpage

%\section{Event reconstruction} 

Event reconstruction algorithms are applied to all of the FC events to
determine the event vertex location, find Cherenkov ring directions,
assign a particle identification of showering or non-showering to each
ring, assign particle momentum to each ring, and identify muon decay
electrons that follow the primary event by microseconds. The same
reconstruction algorithms are applied to both the observed data and
the MCs. The algorithms are the same as for atmospheric neutrino
analyses, with the details of their algorithms described in
\cite{MS:nim}.

\begin{figure}[htbp]
\begin{center}
\includegraphics[width=.9\linewidth]{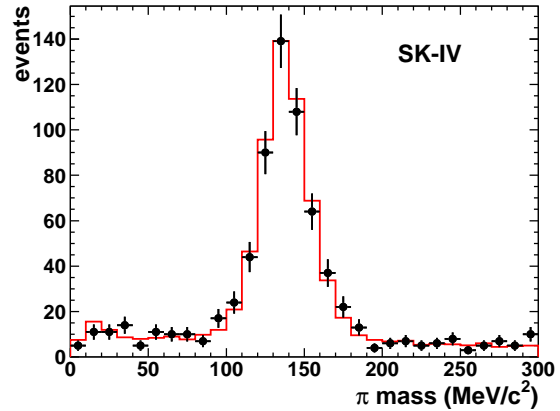}
\end{center}
\caption{ \protect \small Reconstructed invariant mass distribution of
  the atmospheric-neutrino-induced $\pi^0$ events in the observed data
  (black dot) and the atmospheric neutrino MC samples (red histogram)
  used in the absolute energy scale error estimation in SK-IV.  MC
  events are normalized by the livetime of the observed data (1631
  days).  }
\label{fig:sk4pi0}
\end{figure}

One of the most important event reconstructions for nucleon decay
searches is the energy scale, because we distinguish the nucleon decay
events from the atmospheric neutrino background events using their
invariant mass and momentum.  The energy scale uncertainty is
estimated by taking the absolute scale difference between control
sample data and MC as well as the time variation of control sample
data for each SK period. For the four SK periods, the energy scale
uncertainties are 1.1\%, 1.7\%, 2.7\%, and 2.1\% for SK-I, SK-II,
SK-III, and SK-IV, respectively. The uncertainties from SK-I to SK-III
are same as in a previous publication~\cite{Regis:PDK} and details of
the energy reconstruction and the error estimation can be found there.
Figure~\ref{fig:sk4pi0} shows the reconstructed $\pi^0$ mass
distributions used in the absolute energy scale error estimation of
SK-IV.  Good agreement is observed between data and MC in
reconstruction of the $\pi^0$ mass.
Zenith angle-dependent non-uniformities of the energy scale are 0.6\%, 0.6\%, 1.3\%, and 0.7\% for SK-I, SK-II, SK-III, and SK-IV, respectively. These energy scale errors are used in the systematic error estimations described in Section~\ref{sec:syserr}.

The reconstructed total momentum, $P_\textrm{tot}$, the total energy
$E_\textrm{tot}$, and the invariant mass $M_\textrm{tot}$ are defined
as:
\begin{eqnarray}
P_\textrm{tot} &=& \left|\sum_i^{\textrm{all}} \vec{p}_i\right|,\\
E_\textrm{tot} &=& \sum_i^{\textrm{all}}
\sqrt{\left|\vec{p}_i\right|^2+m_i^2},\\
M_\textrm{tot} &=& \sqrt{E_\textrm{tot}^2- P_\textrm{tot}^2},
\end{eqnarray}
where $\vec{p}_i$ is the momentum of each Cherenkov ring and $m_i$ is
the mass of a particle.

The meson mass is reconstructed in a similar way by summing up the
momenta and energies of the secondary particles from the meson decay.
Under the assumption of a particular meson decay, the meson mass is
calculated for all possible combinations of particle type assignment
and the best combination in which the reconstructed mass is the
closest to the expected mass is selected.

\begin{figure}[htbp]
\begin{center}
\includegraphics[width=.9\linewidth]{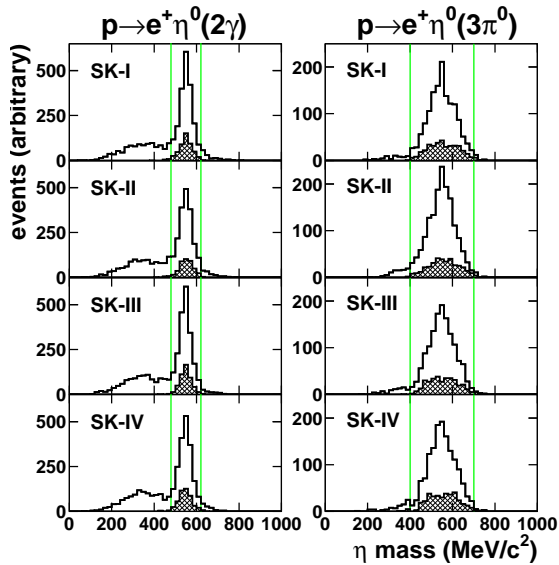}
\end{center}
\caption{ \protect \small The reconstructed $\eta$ mass
  distributions for $p \rightarrow e^+ \eta$ MC events for $\eta
  \rightarrow 2\gamma$ search (left) and $\eta \rightarrow 3\pi^0$
  (right) search for each SK period.
%The open (hatched) histogram shows all (free) proton decay events.
%For the free proton decay MCs, true 2$\gamma$ (3$\pi^0$) are also required for each search.
%For the free proton decay MCs, true 2$\gamma$ and 3$\pi^0$ decays are shown for each search, respectively.
The histograms show all simulated $p \rightarrow e^+ \eta$ events.
For the left plots the eta meson is reconstructed under the assumption of $\eta\rightarrow 2\gamma$, and for the right plots the eta meson is reconstructed under the assumption of $\eta\rightarrow 3\pi^0$.
The hatched histograms show, for free proton decays, the true $\eta\rightarrow 2\gamma$ events (left) and true $\eta\rightarrow 3\pi^0$ events (right).
The event selections (A1-A2) of Section~\ref{sec:p2leta} are applied.
The inside of the two vertical green lines corresponds to the signal region.
}
\label{fig:etamass}
\end{figure}

For example, Fig.~\ref{fig:etamass} shows the reconstructed $\eta$
mass distributions for the $p \rightarrow e^+ \eta$ MC events for
each SK period.
The $\eta$ mass is reconstructed well. 
%All decay branches of the $\eta$ meson are filled in the histograms for the proton decay MC and the other peak outside of the signal window are events from the other decay branches.
The open histogram contains all the relevant $\eta$ decay channels
associated with the signal MC. For example, the broad peak between
200-400~MeV/c$^2$ in the left plots is from all the $\eta$ decays
except for $\eta\rightarrow 2\gamma$.  The mass resolution is observed
to be worse for the 3$\pi^0$ search, since there are six true gamma
rays in the final state but only up to four or five Cherenkov rings are
reconstructed; the ring-counting algorithm terminates at five
rings. For both meson decay searches, there is no significant
difference of the mass reconstruction among the four SK periods.
Details of the comparison of event reconstructions between SK-I and SK-II
are found in the previous publication~\cite{Nishino:2012bnw}.

%All the event reconstruction parameters used in the event selections are checked using the nucleon decay and the atmospheric neutrino MC events for each nucleon decay search and each SK period before opening the data.

%\section{Nucleon decay analysis}

%\subsection{Event selection}\label{sec:selections}
\section{Event selection}\label{sec:selections}

The event selections were already optimized in the previous
analysis~\cite{Nishino:2012bnw} for each nucleon decay mode to
maximize the lifetime sensitivity and are exactly the same in this
analysis except when the two-box total momentum separation and the
SK-IV neutron tagging are used.

The event selections for each nucleon decay ($l^+$ = $e^+$, $\mu^+$) search are described in the following subsections. The signal selection efficiency and the number of expected atmospheric neutrino backgrounds, as well as the observed data for each nucleon decay search and each SK period, are summarized in Table~\ref{tab:summaryall}.
%The comparison with the previous analysis~\cite{Nishino:2012bnw} in SK-I and SK-II is shown in Table~\ref{tab:summary12}.

\subsection{$p\rightarrow l^+  \eta$ search}\label{sec:p2leta}

The $\eta$ meson, with a mass of 548~MeV/c$^2$ and a lifetime of
5$\times$10$^{-19}$ seconds, has three dominant decay modes. Two of
them, $\eta \rightarrow 2\gamma$ (branching ratio = 39\%) and $\eta
\rightarrow 3\pi^0$ (branching ratio = 33\%), are analyzed.  The $\eta
\rightarrow \pi^+\pi^-\pi^0$ (branching ratio = 23\%) is not used
because the branching ratio is smaller and the signal selection efficiency
is worse~\cite{Nishino:2012bnw}.

\vspace{0.1in}
\paragraph{$\eta \rightarrow 2\gamma$ search}
\vspace{0.1in}

~\newline

The following event selection is used:

\begin{itemize}
\item[(A1)] the number of Cherenkov rings is three,
\item[(A2)] all the rings are shower type rings for $p \rightarrow e^+
  \eta$ and one of the rings is a non-shower type ring for $p
  \rightarrow \mu^+ \eta$,
\item[(A3)] the invariant mass is between 480 and 620~MeV/c$^2$,
\item[(A4)] the number of Michel electrons is 0 for $p \rightarrow e^+
  \eta$ or 1 for $p \rightarrow \mu^+ \eta$,
\item[(A5)] the total momentum is less than 250~MeV/c and the
  invariant mass is between 800 and 1050~MeV/c$^2$,
\item[(A6)] the total momentum is greater than or (A7) less than
  100~MeV/c (total momentum separation),
\item[(A8)] the number of neutrons is 0 for the upper or (A9) lower
  total momentum region in SK-IV.
\end{itemize}

After the selections (A1-A5), the total momentum separation (A6-A7)
into two search boxes is implemented, since the following three extra
criteria are satisfied in the lower-momentum signal box: (1) the
expected number of the background events is negligibly small
($\ll$0.1~events for each SK period), (2) the signal efficiency is
sufficiently high ($>$1\%), (3) most of the free proton decays are in
this region.

\begin{figure*}[htbp]
\begin{center}
\includegraphics[width=150mm]{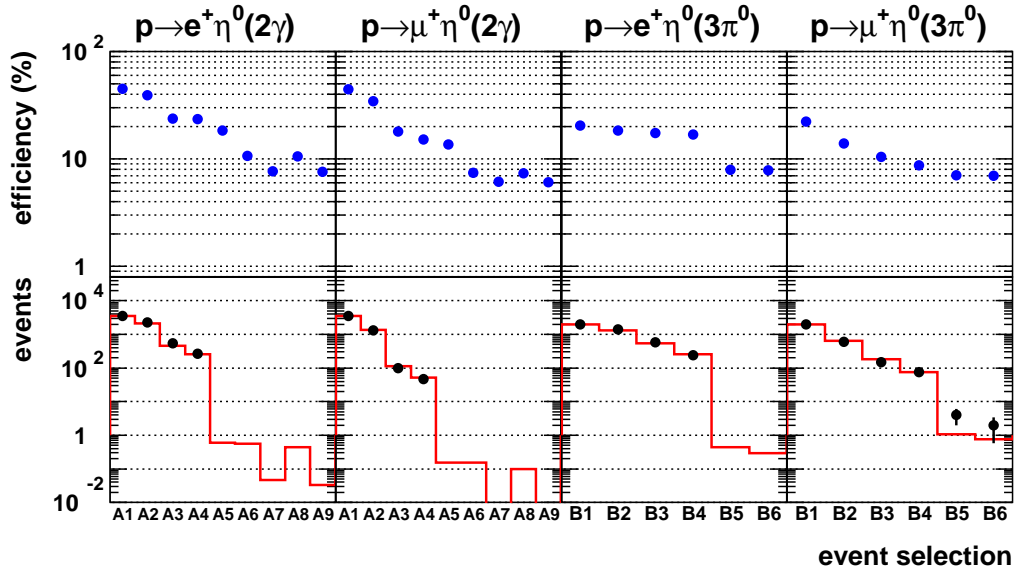}
\end{center}
\caption{ \protect \small The signal efficiencies (upper) and the
  number of expected backgrounds (lower, red histogram) and data
  candidates (lower, black dots) for $p \rightarrow e^+ \eta$ ($\eta
  \rightarrow 2\gamma$), $p \rightarrow \mu^+ \eta$ ($\eta \rightarrow
  2\gamma$), $p \rightarrow e^+ \eta$ ($\eta \rightarrow 3\pi^0$), and
  $p \rightarrow \mu^+ \eta$ ($\eta \rightarrow 3\pi^0$) searches from
  left to right.  The results from SK-I to SK-IV are combined.  The
  event selection is defined in the text.  }
\label{fig:p2leta_nevts}
\end{figure*}

%A minor bug in the $\eta$ mass calculation [Nishino] is fixed in this analysis for $p \rightarrow \mu^+ \eta$, $\eta \rightarrow 3\pi^0$ search.
The signal selection efficiency and the number of expected background
events at each event selection step are shown in
Fig.~\ref{fig:p2leta_nevts}.
The event selection by total momentum and invariant mass at selection (A5) significantly reduces the atmospheric neutrino background. 
The total expected number of the background is further reduced and becomes negligibly small, $\sim$0.03 ($<$0.01) integrated over all the SK periods for $p \rightarrow e^+ \eta$ ($p \rightarrow \mu^+ \eta$) search (Table~\ref{tab:summaryall} for detail), in the lower signal box at selection A7. The signal efficiency is still sufficiently high ($>$1\%) in the lower box.

\begin{figure*}[htbp]
\begin{center}
\includegraphics[width=150mm]{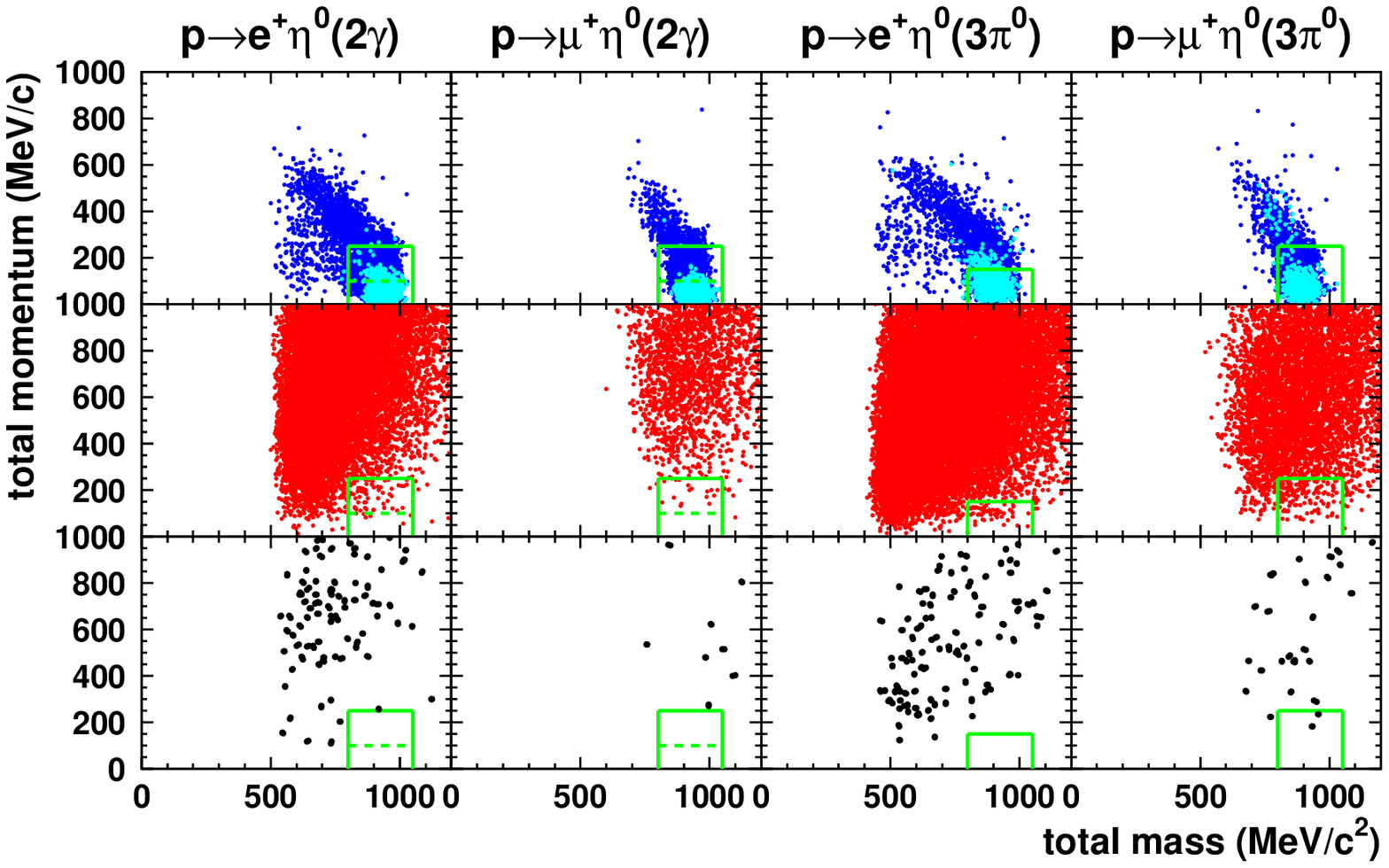}
\end{center}
\caption{ \protect \small Total invariant mass and total momentum for
  $p \rightarrow l^+ \eta$ MC (top, cyan for free proton and blue
  for bound proton), atmospheric neutrino background MC corresponding
  to about 2000 years live time of SK (middle), and data (bottom) for $p
  \rightarrow e^+ \eta$ ($\eta \rightarrow 2\gamma$), $p
  \rightarrow \mu^+ \eta$ ($\eta \rightarrow 2\gamma$), $p
  \rightarrow e^+ \eta$ ($\eta \rightarrow 3\pi^0$), and $p
  \rightarrow \mu^+ \eta$ ($\eta \rightarrow 3\pi^0$) searches
  from left to right.  All the event selections except (A5-A7) or (B5)
  in Section~\ref{sec:p2leta} are applied.  The results from SK-I to
  SK-IV are combined.  The signal box is shown as a green box.  For
  the $p \rightarrow l^+ \eta$ (2$\gamma$) searches, the total
  momentum is separated at the horizontal dashed line in the signal
  box.  }
\label{fig:p2leta_2d}
\end{figure*}

\begin{figure*}[htbp]
\begin{center}
\includegraphics[width=150mm]{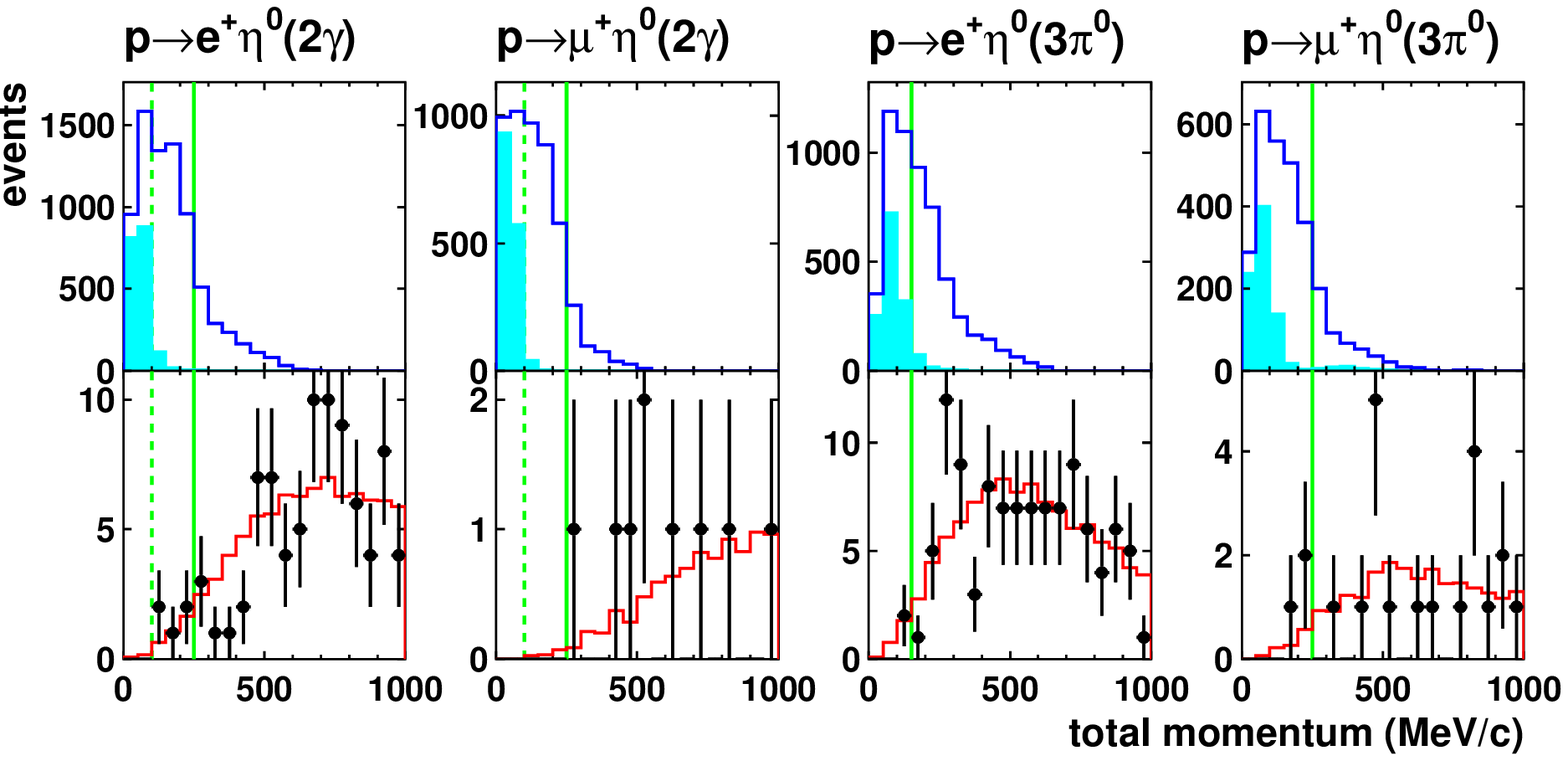}
\end{center}
\caption{ \protect \small Total momentum for $p \rightarrow l^+
  \eta$ MC (upper, open histogram for bounded protons and shaded
  histogram for free protons), atmospheric neutrino MC (lower, red
  histogram), and data (lower, black circles) for $p \rightarrow e^+
  \eta$ ($\eta \rightarrow 2\gamma$), $p \rightarrow \mu^+ \eta$
  ($\eta \rightarrow 2\gamma$), $p \rightarrow e^+ \eta$ ($\eta
  \rightarrow 3\pi^0$), and $p \rightarrow \mu^+ \eta$ ($\eta
  \rightarrow 3\pi^0$) searches from left to right.  All the event
  selections except (A5-A7) or (B5) in Section~\ref{sec:p2leta} are
  applied.  The atmospheric neutrino MC is normalized to data by area.
  Results from SK-I to SK-IV are combined.  The signal region
  corresponds to left side from the vertical green solid line.  The
  vertical dashed line for $p \rightarrow l^+ \eta$ (2$\gamma$)
  corresponds to the total momentum separation cut value.  Total
  momenta for the $p \rightarrow \mu^+ \eta$ ($\eta \rightarrow
  3\pi^0$) data candidates are 188~MeV/c and 239~MeV/c.}
\label{fig:p2leta_ptot}
\end{figure*}

The total mass and total momentum distributions are shown in Fig.~\ref{fig:p2leta_2d} and Fig.~\ref{fig:p2leta_ptot}.
Most of the free protons decay in the lower box (total momentum $<$ 100~MeV/c) for the $\eta\rightarrow 2\gamma$ searches shown in the upper-left two plots and the data and the atmospheric neutrino MC agree with each other in all the lower plots in Fig.~\ref{fig:p2leta_ptot}.
%
%Both total mass and momentum are reconstructed well. Most of the free protons decay in the lower signal box. The same criteria are applied for all the nucleon decay searches and only this search meets with the extra criteria to implement the total momentum separation.

\vspace{0.1in}
\paragraph{$\eta \rightarrow 3\pi^0$ search}
\vspace{0.1in}

~\newline

The following event selection is used:

\begin{itemize}
\item[(B1)] the number of Cherenkov rings is four or five,
\item[(B2)] all the rings are shower type rings for $p \rightarrow e^+ \eta$ and one of the rings is a non-shower type ring for $p \rightarrow \mu^+ \eta$,
\item[(B3)] the $\eta$ mass is between 400 and 700~MeV/c$^2$,
\item[(B4)] the number of Michel electrons is 0 for $p \rightarrow e^+
  \eta$ and 1 for $p \rightarrow \mu^+ \eta$,
\item[(B5)] the total momentum is less than 150~MeV/c for $p
  \rightarrow e^+ \eta$ and less than 250~MeV/c for $p \rightarrow
  \mu^+ \eta$; the total invariant mass is between 800 and
  1050~MeV/c$^2$,
\item[(B6)] the number of neutrons is 0 in SK-IV.
\end{itemize}
%Difference of the total momentum cut values between $e^+$ and $\mu^+$ at selection 5 comes from difference of the number of remaining background events.

The signal selection efficiency and the number of expected background events at each event selection step are shown in Fig.~\ref{fig:p2leta_nevts}. 
The total mass and total momentum distributions are shown in Fig.~\ref{fig:p2leta_2d} and Fig.~\ref{fig:p2leta_ptot}.
Unlike the $\eta \rightarrow 2\gamma$ mode, a significant fraction of
the events have reconstructed total momentum greater than 100~MeV/c
even for free proton decay. Therefore we do not implement the two-box
total momentum separation.

After the full event selection in the $p \rightarrow \mu^+ \eta$ search has been applied, there are two data candidates remaining, one in SK-II and another in SK-IV. 
%(The event displays for all the data candidates will be submitted as {\it supplemental material} of PRD.)
%The event displays are shown in Figure~\ref{fig:p2mueta_3pi_ev1}--\ref{fig:p2mueta_3pi_ev2}.
%The events are well reconstructed.
%
The total number of expected background events integrated over all the
meson decay modes and all SK periods is 0.85. The Poisson probability
to observe two or more events is 20.9\%, as summarized in
Table~\ref{tab:summaryall2}.

The data candidate found in SK-II using the current analysis was also
found in the previously published analysis.  The previous work also
found an additional data candidate in SK-II. In that event, a faint
non-shower like ring reconstructed in the previous analysis is not
reconstructed in this work, excluding it from the selection
criteria. Event reconstruction improvements, such as tuning of the
backward charge within the charge separation algorithm among Cherenkov
rings, can cause such changes in the ring counting outcome on an
event-by-event basis.

\subsection{$p\rightarrow l^+ \rho^0$}\label{sec:p2lrho}

The $\rho^0$ meson, with a mass of 775~MeV/c$^2$ and a width of
149~MeV, decays immediately into $\pi^+\pi^-$ with a branching ratio
of nearly 100\%.

~\newline

The following event selection is used for this search:

\begin{itemize}
\item[(C1)] the number of Cherenkov rings is three,
\item[(C2)] one of the rings is a shower type ring for $p \rightarrow e^+ \rho^0$ and all the rings are non-shower type rings for $p \rightarrow \mu^+ \rho^0$,
\item[(C3)] the $\rho^0$ mass is between 600 and 900~MeV/c$^2$,
\item[(C4)] the number of Michel electrons is 0 or 1 for $p
  \rightarrow e^+ \rho^0$ and 1 or 2 for $p \rightarrow \mu^+ \rho^0$,
\item[(C5)] the total momentum is less than 150~MeV/c for $p
  \rightarrow e^+ \rho^0$ and less than 250 MeV/c for $p \rightarrow
  \mu^+ \rho^0$; the total invariant mass is between 800 and
  1050~MeV/c$^2$,
\item[(C6)] the number of neutrons is 0 in SK-IV.
\end{itemize}

\begin{figure}[htbp]
\begin{center}
\includegraphics[width=.9\linewidth]{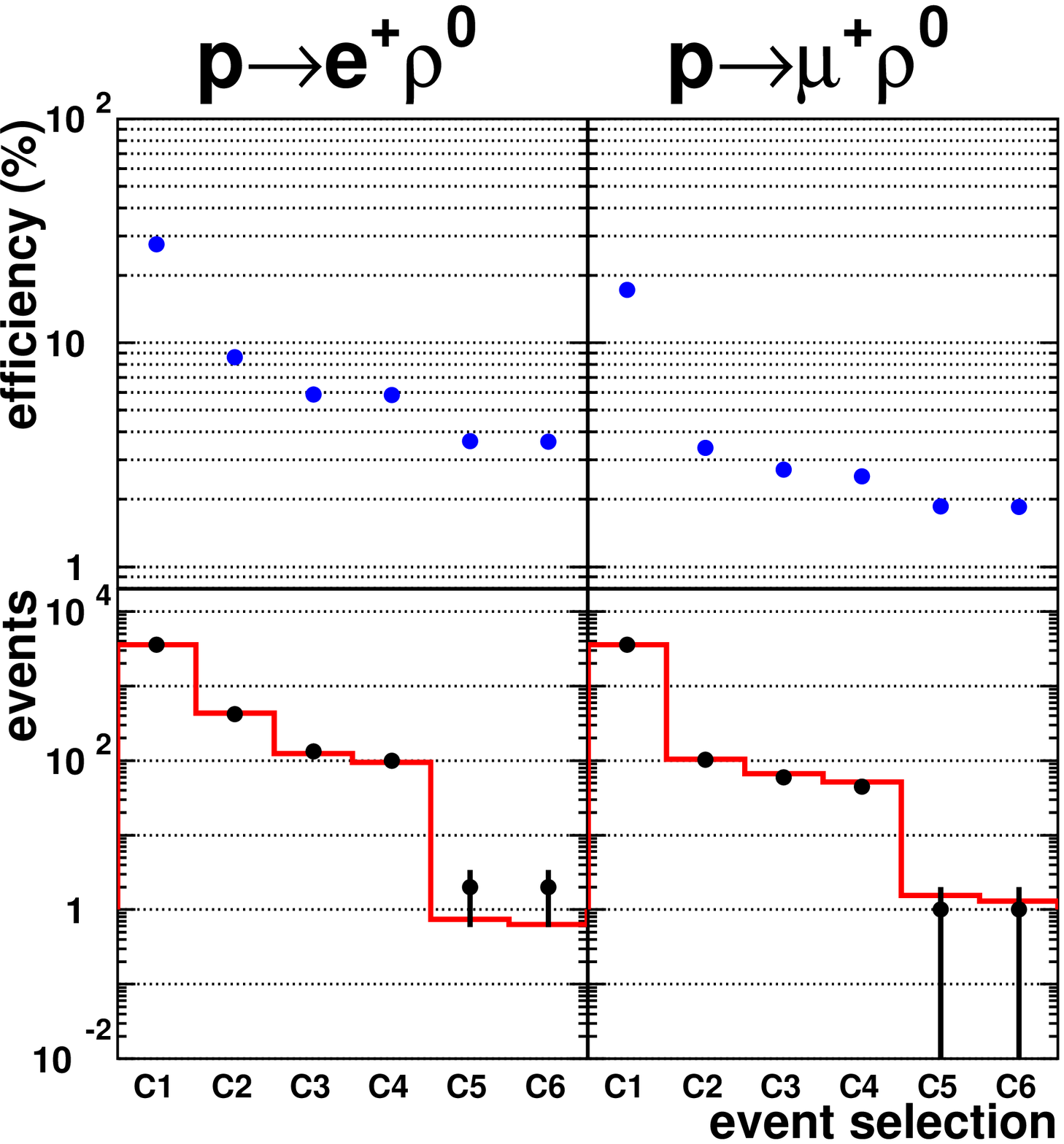}
\end{center}
\caption{ \protect \small
The signal efficiencies (upper) and the number of expected backgrounds (lower, red histogram) and data candidates (lower, black dots) for $p \rightarrow e^+ \rho^0$ (left) and $p \rightarrow \mu^+ \rho^0$ (right) searches.
The results from SK-I to SK-IV are combined.  The event selection is defined in the text.
}
\label{fig:p2lrho_nevts}
\end{figure}

\begin{figure}[htbp]
\begin{center}
\includegraphics[width=.9\linewidth]{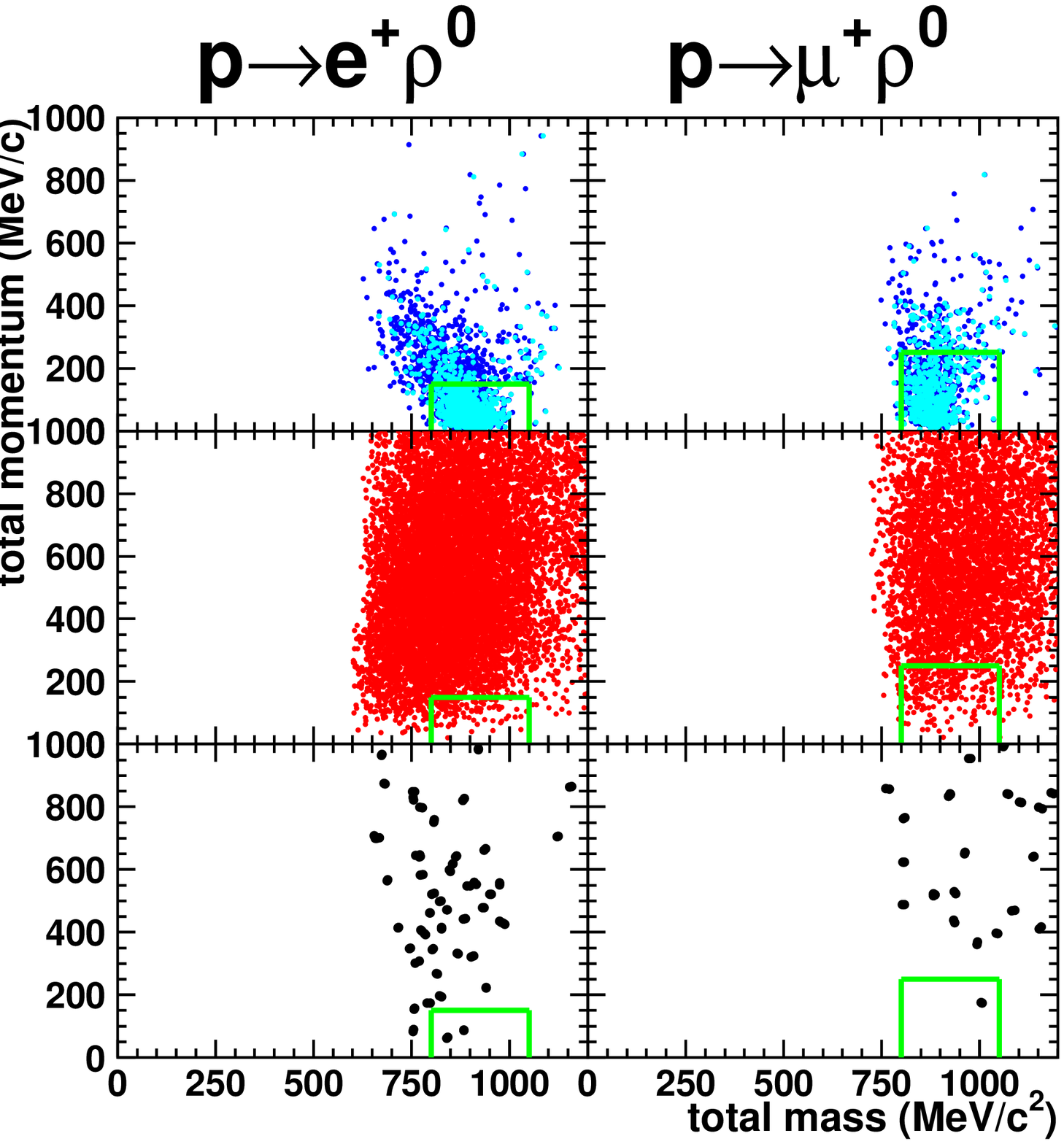}
\end{center}
\caption{ \protect \small Total invariant mass and total momentum for
  $p \rightarrow l^+ \rho^0$ MC (top, cyan for free proton and blue
  for bound proton), atmospheric neutrino background MC corresponding
  to about 2000 years live time of SK (middle), and data (bottom) for $p
  \rightarrow e^+ \rho^0$ (left) and $p \rightarrow \mu^+ \rho^0$
  (right) searches.  All the event selections except (C5) in
  Section~\ref{sec:p2lrho} are applied.  The results from SK-I to
  SK-IV are combined.  The signal box is shown as a green box.  }
\label{fig:p2lrho_2d}
\end{figure}

\begin{figure}[htbp]
\begin{center}
\includegraphics[width=.9\linewidth]{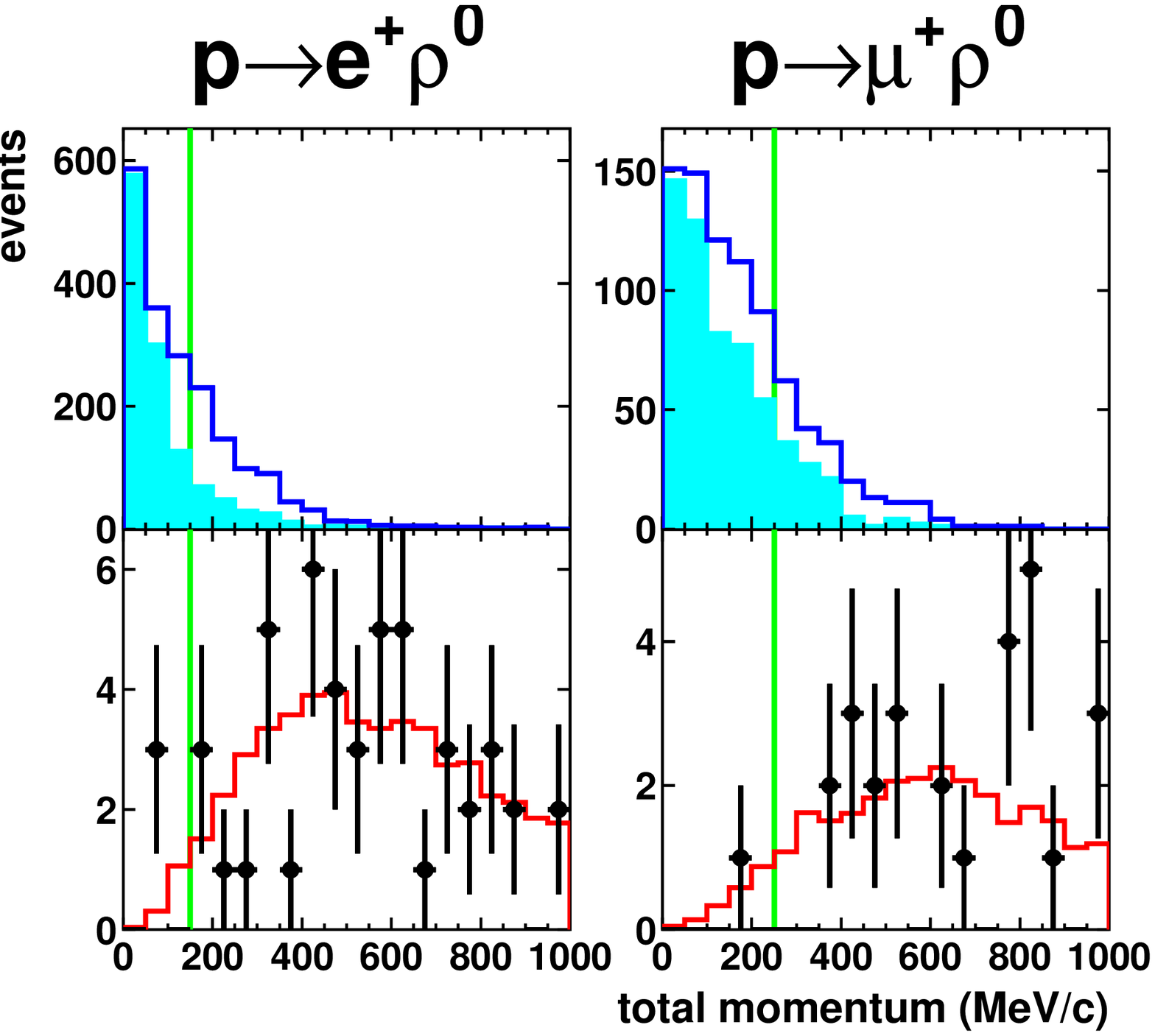}
\end{center}
\caption{ \protect \small
Total momentum for $p \rightarrow l^+ \rho^0$ MC (upper, open histogram for bounded protons and shaded histogram for free protons), atmospheric neutrino MC (lower, red histogram), and data (lower, black circles) for $p \rightarrow e^+ \rho^0$ (left) and $p \rightarrow \mu^+ \rho^0$ (right) searches.
All the event selections except (C5) in Section~\ref{sec:p2lrho} are applied.
The atmospheric neutrino MC is normalized to data by area.
Results from SK-I to SK-IV are combined.
The signal region corresponds to left side from the vertical green line.
Total momenta are 61~MeV/c and 90~MeV/c for the $p \rightarrow e^+ \rho^0$ data candidates and 179~MeV/c for the $p \rightarrow \mu^+ \rho^0$ data candidate.
}
\label{fig:p2lrho_ptot}
\end{figure}

Figure~\ref{fig:p2lrho_nevts} shows the signal selection efficiency
and the number of expected background events at each event selection
step.
The total invariant mass and momentum distributions are shown in
Figs.~\ref{fig:p2lrho_2d} and \ref{fig:p2lrho_ptot}.

There are two data candidates in SK-IV for $p \rightarrow e^+ \rho^0$
search and one candidate in SK-I for $p \rightarrow \mu^+ \rho^0$
search. The SK-I event was also selected in the analysis of the
previous publication. The total number of expected background events
is 0.64 for $p \rightarrow e^+ \rho^0$ and 1.30 for $p
\rightarrow\mu^+ \rho^0$ searches.
%The event displays are shown in Figure~\ref{fig:p2erho_ev1}--\ref{fig:p2murho_ev1}. 

\subsection{$p\rightarrow l^+ \omega$}\label{sec:p2lomega}

The $\omega$ meson has a mass of 783~MeV/c$^2$ and immediately
decays due to a width of 8.5 MeV. Two decay modes are analyzed: one is
the $\omega \rightarrow \pi^+ \pi^- \pi^0$ mode (branching ratio =
89\%), and the other is the $\omega \rightarrow \pi^0 \gamma$ mode
(branching ratio = 8\%).

\vspace{0.1in}
\paragraph{$\omega \rightarrow \pi^0 \gamma$ search}
\vspace{0.1in}

~\newline

The following event selection is used for this search:

\begin{itemize}
\item[(D1)] the number of Cherenkov rings is 3 or 4 for $p \rightarrow
  e^+ \omega$ and 2 or 3 for $p \rightarrow \mu^+ \omega$,
\item[(D2)] all rings are shower type rings,
\item[(D3)] the $\omega$ mass is between 650 and 900~MeV/c$^2$,
\item[(D4)] the number of Michel electrons is 0 for $p \rightarrow e^+
  \omega$ and 1 for $p \rightarrow \mu^+ \omega$,
\item[(D5)] the total momentum is less than 150~MeV/c for $p
  \rightarrow e^+ \omega$ and less than 200~MeV/c for $p \rightarrow
  \mu^+ \omega$; the invariant mass is between 800 and
  1050~MeV/c$^2$ for $p \rightarrow e^+ \omega$,
\item[(D6)] the number of neutrons is 0 in SK-IV.
\end{itemize}

For the $p \rightarrow \mu^+ \omega$ mode, the muon momentum is lower than the Cherenkov threshold and the muon ring is not observed.

\begin{figure*}[htbp]
\begin{center}
\includegraphics[width=150mm]{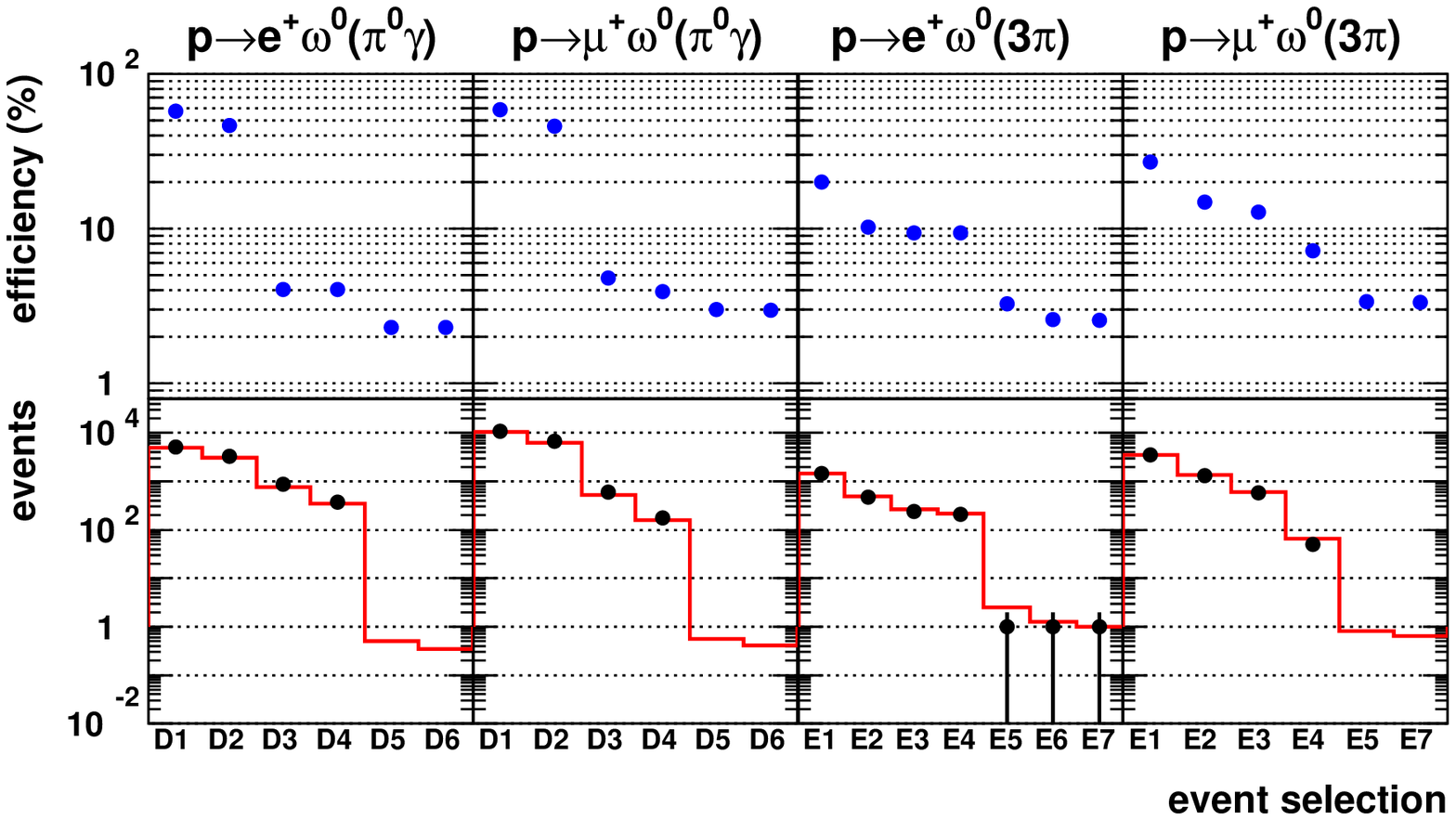}
\end{center}
\caption{ \protect \small
The signal efficiencies (upper) and the number of expected backgrounds (lower, red histogram) and data candidates (lower, black dots) for $p \rightarrow e^+ \omega$ ($\omega \rightarrow \pi^0\gamma$), $p \rightarrow \mu^+ \omega$ ($\omega \rightarrow \pi^0\gamma$), $p \rightarrow e^+ \omega$ ($\omega \rightarrow 3\pi$), and $p \rightarrow \mu^+ \omega$ ($\omega \rightarrow 3\pi$) searches from left to right.
The results from SK-I to SK-IV are combined.  The event selection is defined in the text.
}
\label{fig:p2lomega_nevts}
\end{figure*}

\begin{figure*}[htbp]
\begin{center}
\includegraphics[width=150mm]{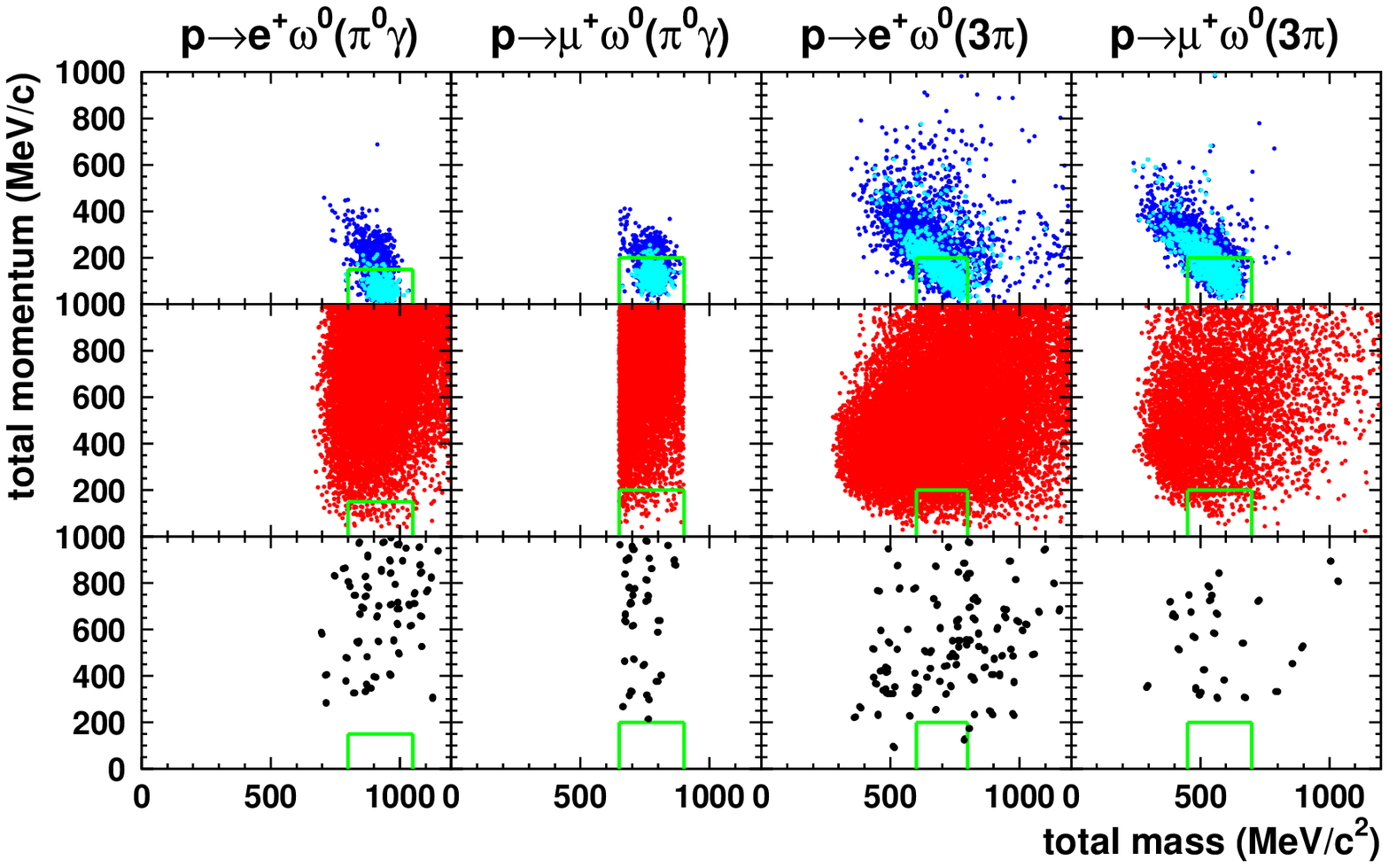}
\end{center}
\caption{ \protect \small Total invariant mass and total momentum for
  $p \rightarrow l^+ \omega$ MC (top, cyan for free proton and blue
  for bound proton), atmospheric neutrino background MC corresponding
  to about 2000 years live time of SK (middle), and data (bottom) for $p
  \rightarrow e^+ \omega$ ($\omega \rightarrow \pi^0\gamma$), $p
  \rightarrow \mu^+ \omega$ ($\omega \rightarrow \pi^0\gamma$), $p
  \rightarrow e^+ \omega$ ($\omega \rightarrow 3\pi^0$), and $p
  \rightarrow \mu^+ \omega$ ($\omega \rightarrow 3\pi^0$) searches
  from left to right.  All the event selections except (D5) or (E5) in
  Section~\ref{sec:p2lomega} are applied.  The results from SK-I to
  SK-IV are combined.  The signal box is shown as a green box.  }
\label{fig:p2lomega_2d}
\end{figure*}

\begin{figure*}[htbp]
\begin{center}
\includegraphics[width=150mm]{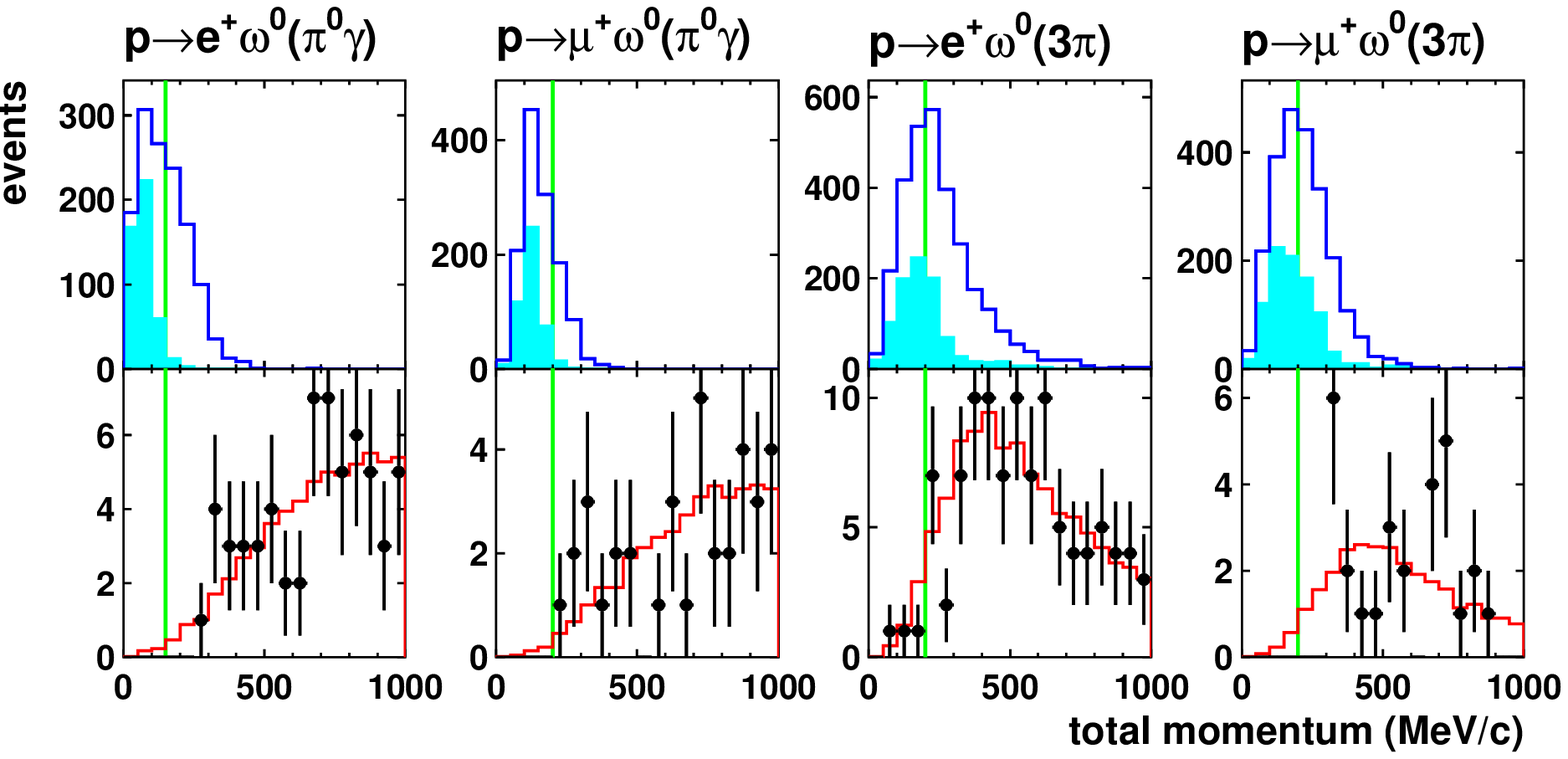}
\end{center}
\caption{ \protect \small
Total momentum for $p \rightarrow l^+ \omega$ MC (upper, open histogram for bounded protons and shaded histogram for free protons), atmospheric neutrino MC (lower, red histogram), and data (lower, black circles) for $p \rightarrow e^+ \omega$ ($\omega \rightarrow \pi^0\gamma$), $p \rightarrow \mu^+ \omega$ ($\omega \rightarrow \pi^0\gamma$), $p \rightarrow e^+ \omega$ ($\omega \rightarrow 3\pi$), and $p \rightarrow \mu^+ \omega$ ($\omega \rightarrow 3\pi$) searches from left to right.
All the event selections except (D5) or (E5) in Section~\ref{sec:p2lomega} are applied.
The atmospheric neutrino MC is normalized to data by area.
Results from SK-I to SK-IV are combined.
The signal region corresponds to left side from the vertical green line.
Total momentum for $p \rightarrow e^+ \omega$ ($\omega \rightarrow 3\pi$) data candidate is 124~MeV/c.
}
\label{fig:p2lomega_ptot}
\end{figure*}

Figure~\ref{fig:p2lomega_nevts} shows the signal selection efficiency and the number of expected background events at each event selection step. 
The total mass and total momentum distributions are shown in Fig.~\ref{fig:p2lomega_2d} and Fig.~\ref{fig:p2lomega_ptot}.

\vspace{0.1in}
\paragraph{$\omega \rightarrow \pi^+ \pi^- \pi^0$ search}
\vspace{0.1in}

~\newline

The following event selection is used for this search:

\begin{itemize}
\item[(E1)] the number of Cherenkov rings is 4 for $p \rightarrow e^+ \omega$ and 3 for $p \rightarrow \mu^+ \omega$,
\item[(E2)] one of the rings is a non-shower ring,
\item[(E3)] the $\pi^0$ mass is between 85 and 185~MeV/c$^2$,
\item[(E4)] the number of Michel electrons is 0 or 1 for $p \rightarrow e^+ \omega$ and 2 for $p \rightarrow \mu^+ \omega$,
\item[(E5)] the total momentum is less than 200~MeV/c, and the total invariant mass is between 600 and 800~MeV/c$^2$ for $p \rightarrow e^+ \omega$ 
and between 450 and 700~MeV/c$^2$ for $p \rightarrow \mu^+ \omega$,
\item[(E6)] the positron momentum is between 100 and 200~MeV/c for $p
  \rightarrow e^+ \omega$,
\item[(E7)] the number of neutrons is 0 in SK-IV.
\end{itemize}

\begin{figure}[htbp]
\begin{center}
\includegraphics[width=.9\linewidth]{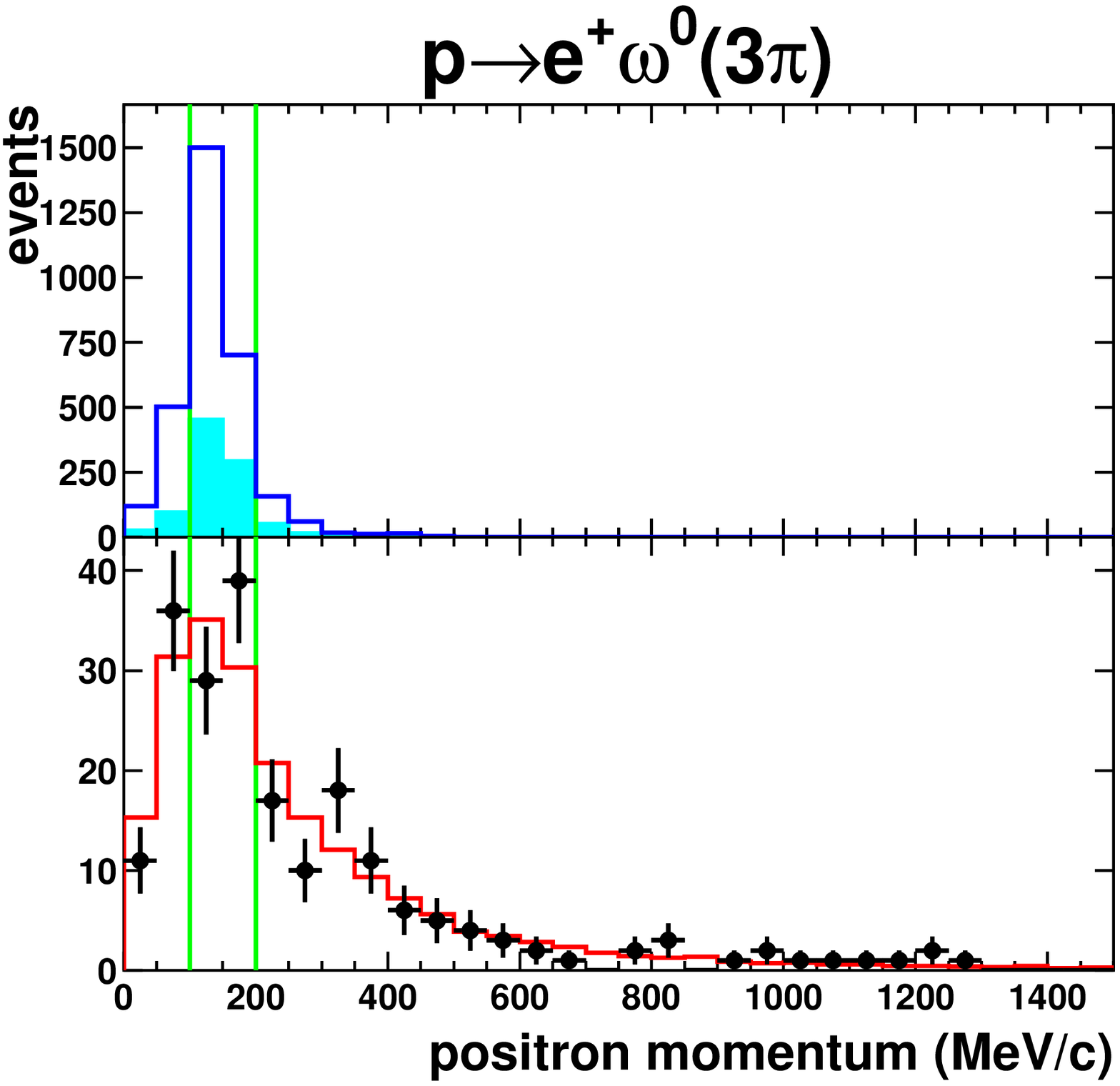}
\end{center}
\caption{ \protect \small
Positron momentum for $p \rightarrow e^+ \omega$ MC (upper, open histogram for bounded protons and shaded histogram for free protons), atmospheric neutrino MC (lower, red histogram), and data (lower, black circles) for $p \rightarrow e^+ \omega$ ($\omega \rightarrow 3\pi$) search.
The event selections (E1-E4) of Section~\ref{sec:p2lomega} are applied.
The atmospheric neutrino MC is normalized to data by area.
Results from SK-I to SK-IV are combined.
Inside of the two vertical green lines corresponds to the signal region.
}
\label{fig:pposi}
\end{figure}

Due to the strong interaction of the charged pions in water, finding
both charged pion rings is difficult. Therefore, we require finding
only one of the two charged pion rings in these searches.  Since one
of the two charged pions is assumed to be invisible in the selection
criteria, the invariant mass for the $\omega$ can not be
reconstructed.  Instead, we require that the reconstructed $\pi^0$
invariant mass is consistent with the $\pi^0$ mass.  In order to
reduce the background even further, the positron momentum is also
required to be consistent with the expected positron momentum for the
$p \rightarrow e^+ \omega$ mode. The positron momentum distribution
is shown in Fig.~\ref{fig:pposi}.

%Figure~\ref{fig:p2lomega_nevts} shows the signal selection efficiency and the number of expected background events at each event selection step. 
%The total mass and total momentum distributions are shown in Figure~\ref{fig:p2lomega_2d} and Figure~\ref{fig:p2lomega_ptot}.

There is one data candidate in SK-I for $p \rightarrow e^+ \omega$ search and the number of total expected background is 1.35. An identical event was selected in the previous analysis. 
%The event display is shown in Figure~\ref{fig:p2eomega_ev1}.

%\clearpage

\subsection{$n\rightarrow l^+ \pi^-$}\label{sec:n2lpi}

~\newline
The following event selection is used for this search:

\begin{itemize}
\item[(F1)] the number of Cherenkov rings is two,
\item[(F2)] one of the rings is a shower type ring for $n \rightarrow
  e^+ \pi^-$ and all the rings are non-shower type rings for $n
  \rightarrow \mu^+ \pi^-$,
\item[(F3)] the number of Michel electrons is 0 for $n \rightarrow
  e^+ \pi^-$ and 1 for $n \rightarrow \mu^+ \pi^-$,
\item[(F4)] the total momentum is less than 250~MeV/c, and the total invariant mass is between 800 and 1050~MeV/c$^2$,
\item[(F5)] the number of neutrons is 0 in SK-IV.
\end{itemize}

In these modes, a Cherenkov ring from the charged pion generated from
the nucleon decay can be directly observed. It is not necessary to
select the events by a meson invariant mass for these modes.

\begin{figure}[htbp]
\begin{center}
\includegraphics[width=.9\linewidth]{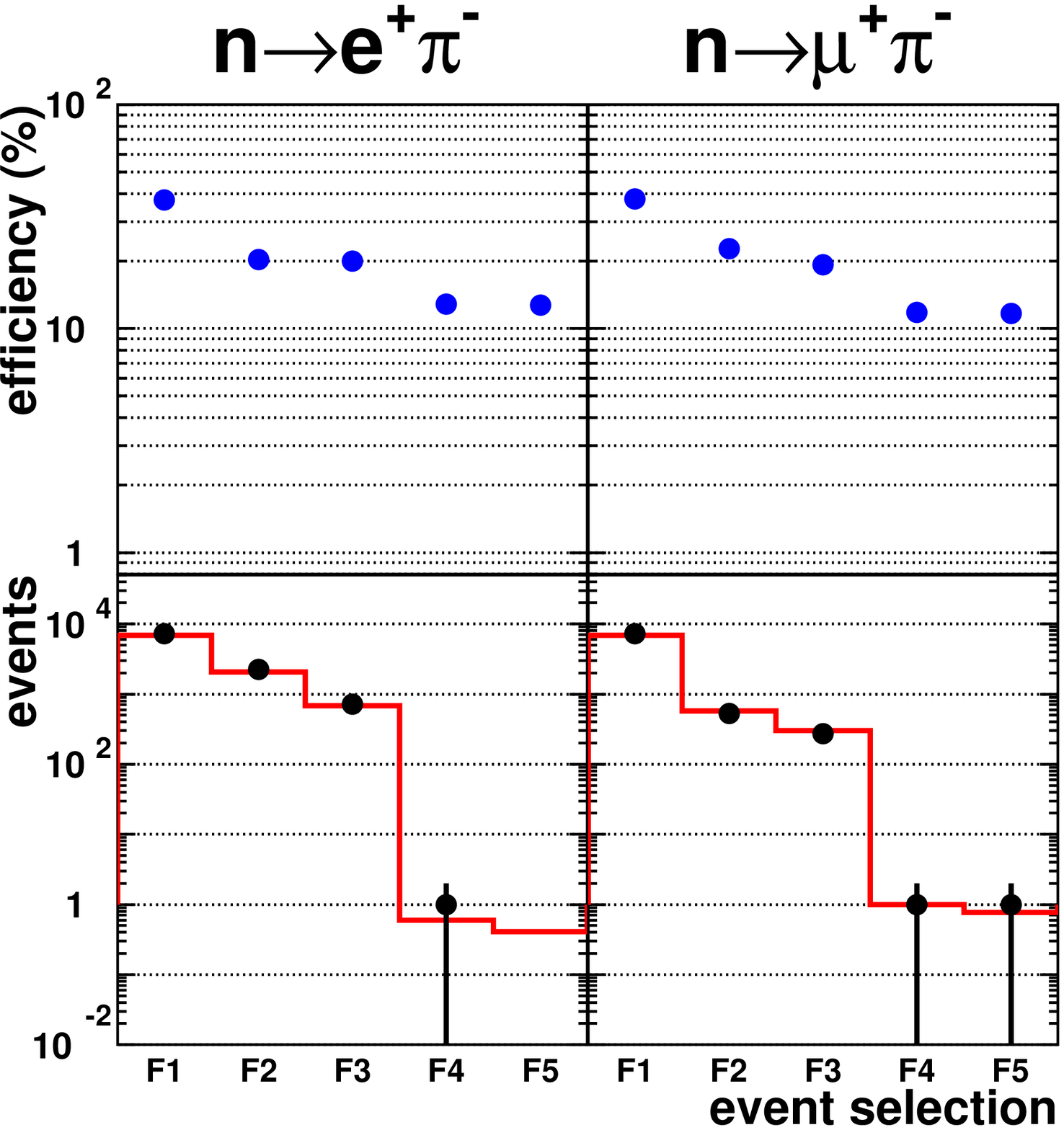}
\end{center}
\caption{ \protect \small
The signal efficiencies (upper) and the number of expected backgrounds (lower, red histogram) and data candidates (lower, black dots) for $n \rightarrow e^+ \pi^-$ (left) and $n \rightarrow \mu^+ \pi^-$ (right) searches.
The results from SK-I to SK-IV are combined.  The event selection is defined in the text.
}
\label{fig:n2lpi_nevts}
\end{figure}

\begin{figure}[htbp]
\begin{center}
\includegraphics[width=.9\linewidth]{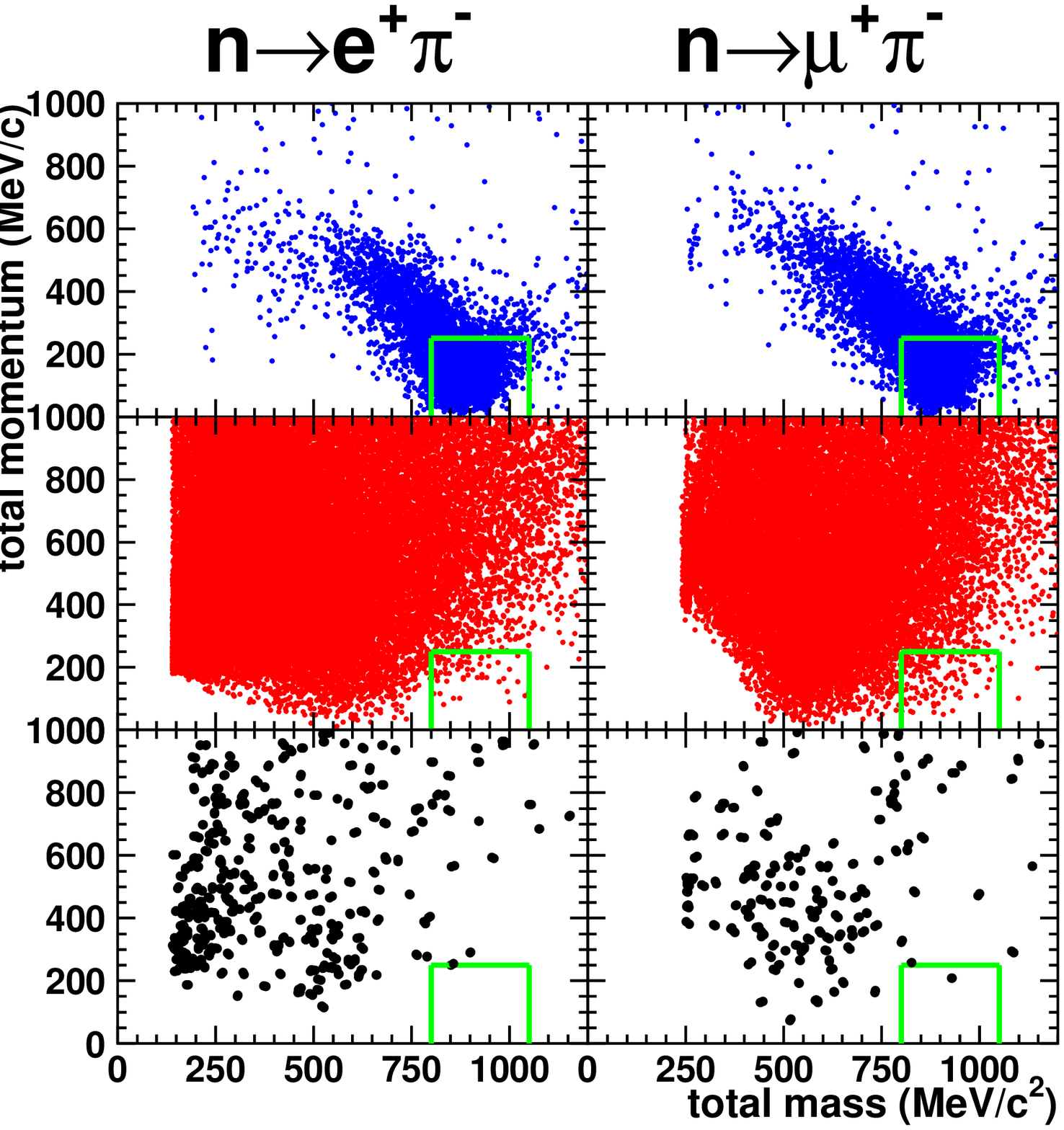}
\end{center}
\caption{ \protect \small Total invariant mass and total momentum for
  $n \rightarrow l^+ \pi^-$ MC (top), atmospheric neutrino background
  MC corresponding to about 2000 years live time of SK (middle), and
  data (bottom) for $n \rightarrow e^+ \pi^-$ (left) and $n
  \rightarrow \mu^+ \pi^-$ (right) searches.  All the event selections
  except (F4) in Section~\ref{sec:n2lpi} are applied.  The results
  from SK-I to SK-IV are combined.  The signal box is shown as a green
  box.  }
\label{fig:n2lpi_2d}
\end{figure}

\begin{figure}[htbp]
\begin{center}
\includegraphics[width=.9\linewidth]{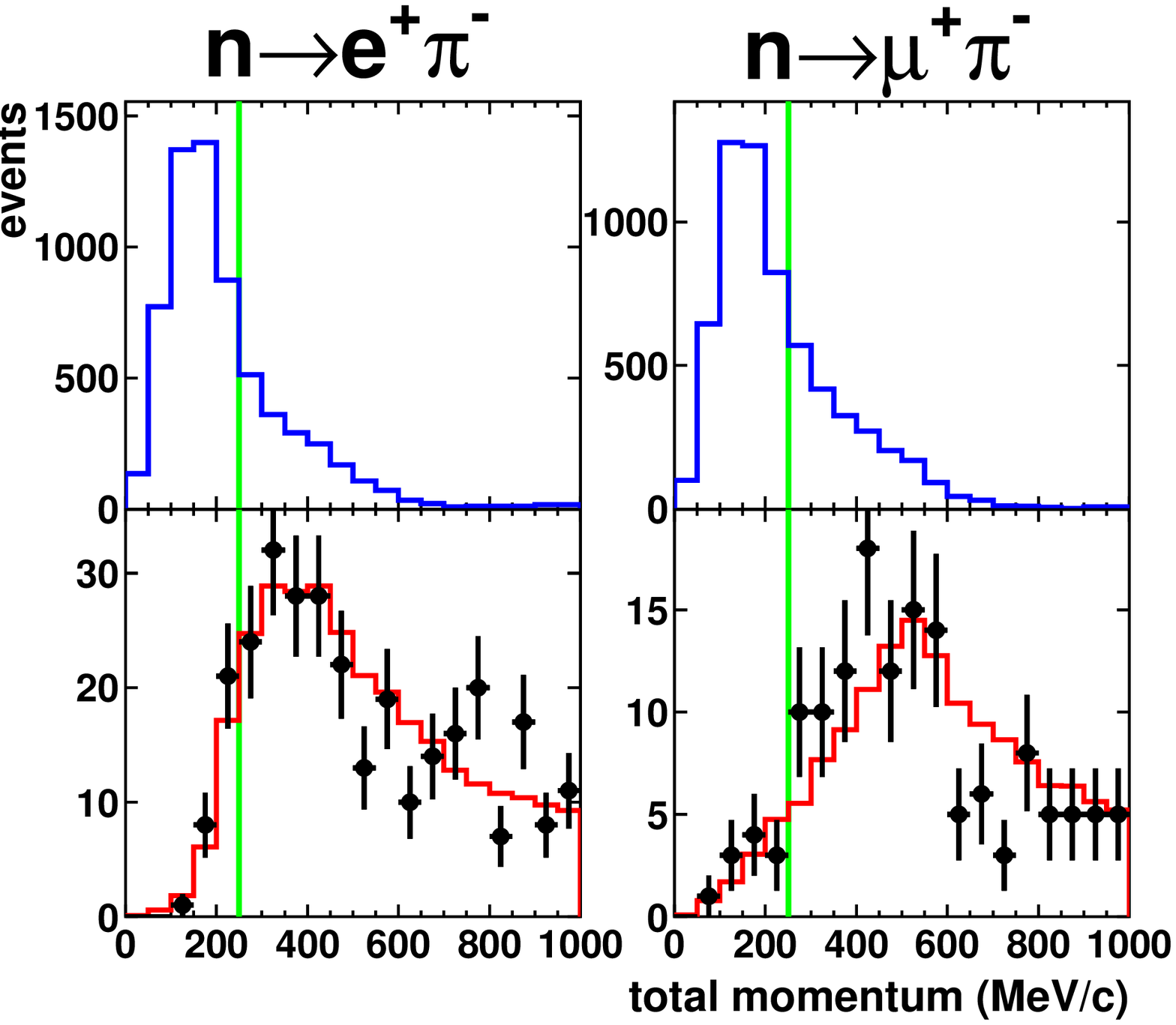}
\end{center}
\caption{ \protect \small
Total momentum for $n \rightarrow l^+ \pi^-$ MC (upper), atmospheric neutrino MC (lower, red histogram), and data (lower, black circles) for $n \rightarrow e^+ \pi^-$ (left) and $n \rightarrow \mu^+ \pi^-$ (right) searches.
All the event selections except (F4) in Section~\ref{sec:n2lpi} are applied.
The atmospheric neutrino MC is normalized to data by area.
Results from SK-I to SK-IV are combined.
The signal region corresponds to left side from the vertical green line.
Total momentum for $n \rightarrow \mu^+ \pi^-$ data candidate is 205~MeV/c. 
}
\label{fig:n2lpi_ptot}
\end{figure}

Figure~\ref{fig:n2lpi_nevts} shows the signal selection efficiency and
the number of expected background events at each event selection step.
The total mass and total momentum distributions are shown in Fig.~\ref{fig:n2lpi_2d} and Fig.~\ref{fig:n2lpi_ptot}.
%A minor bug in the total mass and momentum calculation [Nishino] is fixed in this analysis for $n \rightarrow \mu^+ \pi^-$ search.
%

There is one data candidate in SK-III in $n \rightarrow \mu^+ \pi^-$ search and the number of total expected background events is 0.77. 
%The event display is shown in Figure~\ref{fig:n2mupi_ev1}.
%

In the previous published analysis, there was one data candidate in
SK-I for the $n \rightarrow \mu^+ \pi^-$ search. In both previous and
current analyses, the event vertex is in the fiducial volume, the
number of rings is two, both rings are non-shower types, and the
number of Michel electron is one, passing those selection
criteria. The invariant mass and momentum, ($M_\textrm{tot}$,
$P_\textrm{tot}$) was (809~MeV/c$^2$, 245~MeV/c) in the previous
analysis but changed slightly to (829~MeV/c$^2$, 253~MeV/c) in this
analysis. The $P_\textrm{tot}$ is near the signal box but above the
cut threshold (250~MeV/c) and this event fails final selection for
this analysis.

\subsection{$n\rightarrow l^+ \rho^-$}\label{sec:n2lrho}

The $\rho^-$ meson has a mass of 770~MeV/c$^2$ and
a width of 150~MeV, immediately decaying into $\pi^-\pi^0$ 
with a branching ratio of nearly 100\%.

~\newline
The following event selection is used for this search:

\begin{itemize}
\item[(G1)] the number of Cherenkov rings is 4 for $n
  \rightarrow e^+ \rho^-$ and 3 for $n \rightarrow \mu^+ \rho^-$,
\item[(G2)] one of the rings is a non-shower ring,
\item[(G3)] the $\rho^-$ mass is between 600 and 900~MeV/c$^2$,
\item[(G4)] the $\pi^0$ mass is between 85 and 185~MeV/c$^2$,
\item[(G5)] the number of Michel electrons is 0 for $n \rightarrow e^+
  \rho^-$ and 1 for $n \rightarrow \mu^+ \rho^-$,
\item[(G6)] the total momentum is less than 250~MeV/c for $n
  \rightarrow e^+ \rho^-$ and less than 150~MeV/c for $n \rightarrow
  \mu^+ \rho^-$; the total invariant mass is between 800 and
  1050~MeV/c$^2$ for $n \rightarrow e^+ \rho^-$,
\item[(G7)] the number of neutrons is 0 in SK-IV.
\end{itemize}

The tighter total momentum cut of $P_\textrm{tot} <$ 150~MeV/c is
applied to reduce the background for the $n \rightarrow \mu^+ \rho^-$
mode because the nucleon invariant mass cannot be reconstructed for
the mode due to the invisible muon.

\begin{figure}[htbp]
\begin{center}
\includegraphics[width=.9\linewidth]{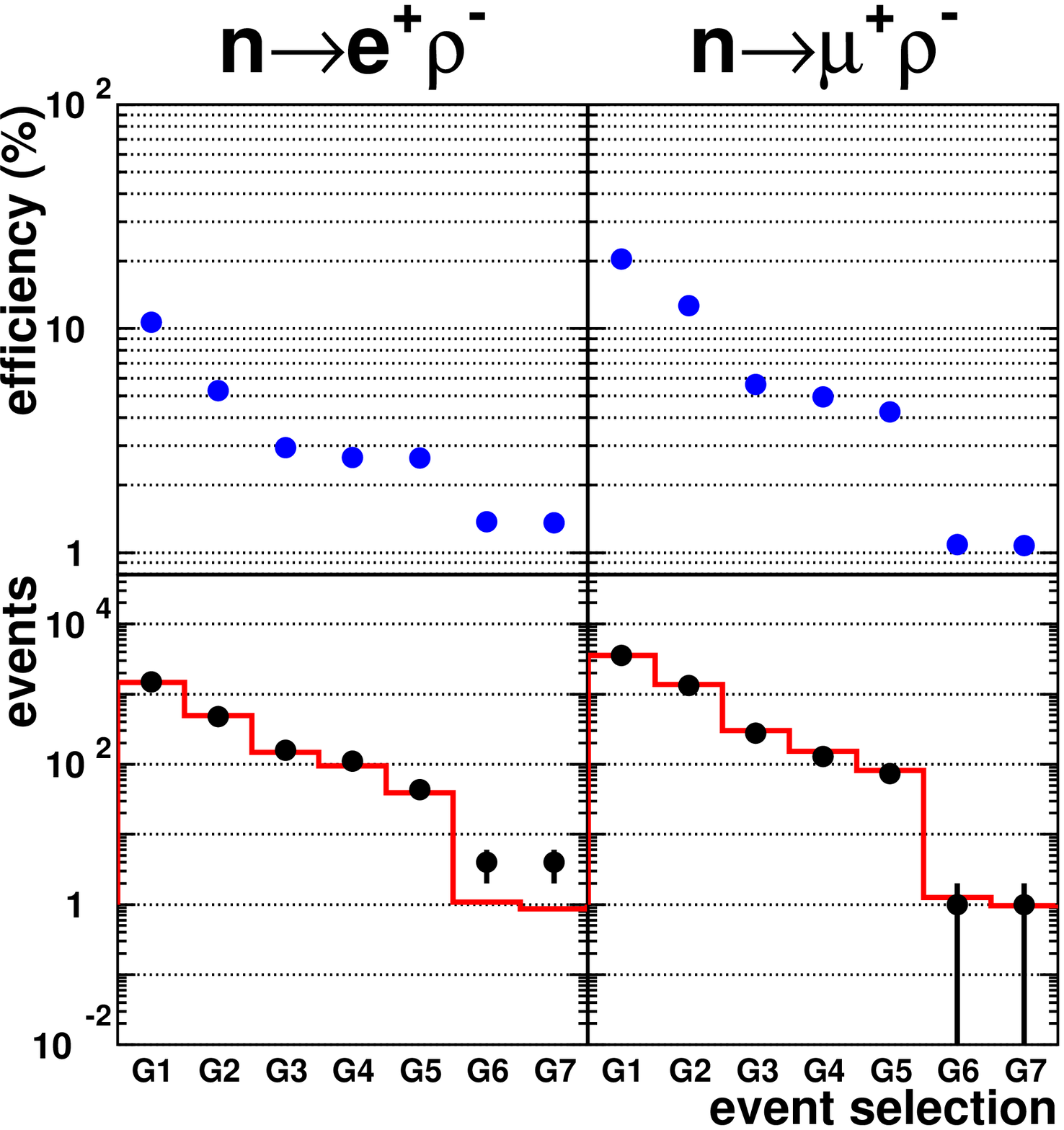}
\end{center}
\caption{ \protect \small
The signal efficiencies (upper) and the number of expected backgrounds (lower, red histogram) and data candidates (lower, black dots) for $n \rightarrow e^+ \rho^-$ (left) and $n \rightarrow \mu^+ \rho^-$ (right).
The results from SK-I to SK-IV are combined.  The event selection is defined in the text.
}
\label{fig:n2lrho_nevts}
\end{figure}

\begin{figure}[htbp]
\begin{center}
\includegraphics[width=.9\linewidth]{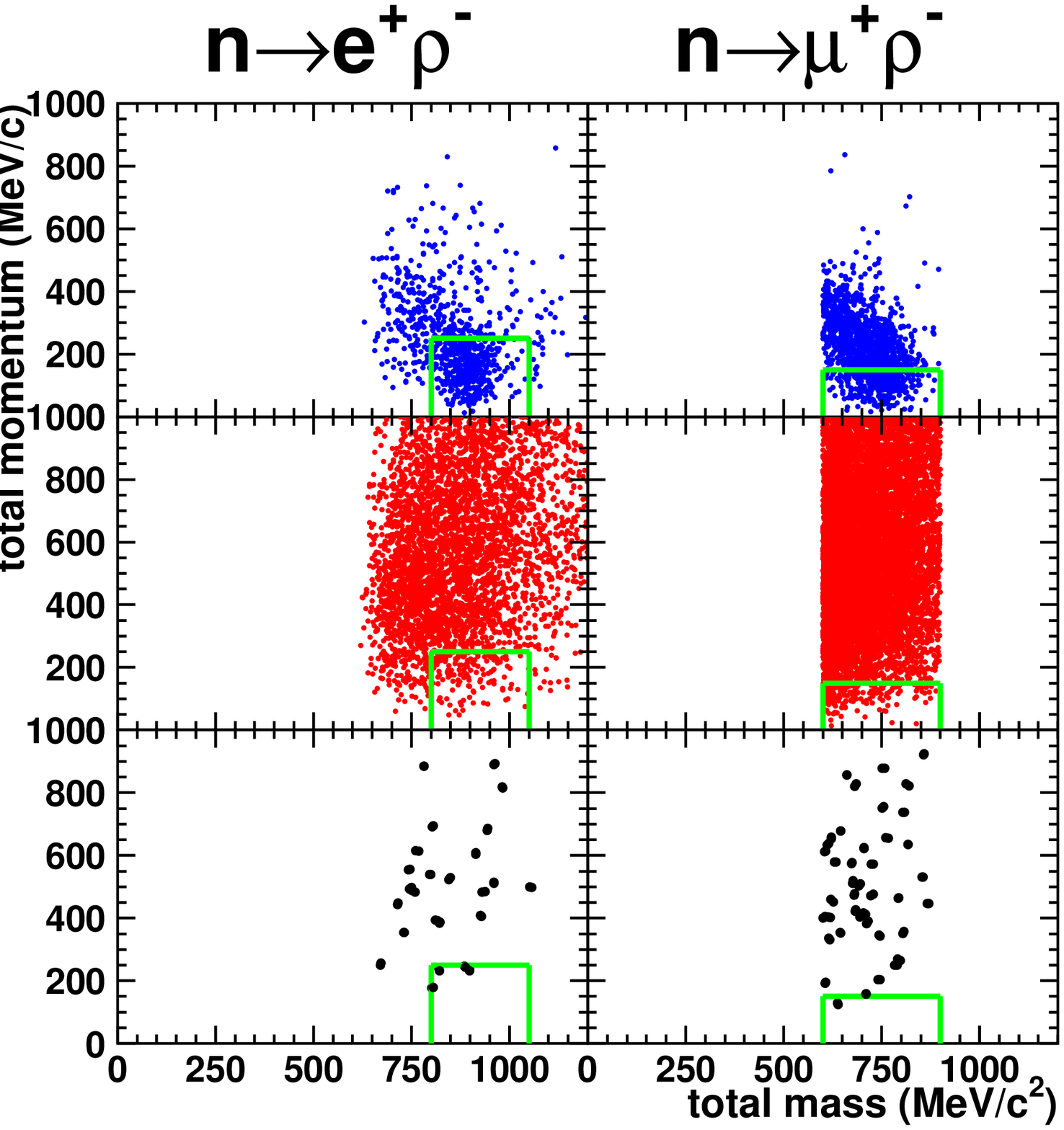}
\end{center}
\caption{ \protect \small Total invariant mass and total momentum for
  $n \rightarrow l^+ \rho^-$ MC (top), atmospheric neutrino background
  MC corresponding to about 2000 years live time of SK (middle), and
  data (bottom) for $n \rightarrow e^+ \rho^-$ (left) and $n
  \rightarrow \mu^+ \rho^-$ (right) searches.  All the event
  selections except (G6) in Section~\ref{sec:n2lrho} are applied.  The
  results from SK-I to SK-IV are combined.  For the $n \rightarrow e^+
  \rho^-$ search, the signal box is shown as a green box.  For the $n
  \rightarrow \mu^+ \rho^-$ search, no total mass cut is applied in
  the analysis and the sharp cutoffs on the total mass correspond to
  the $\rho^-$ mass cut threshold.  }
\label{fig:n2lrho_2d}
\end{figure}

\begin{figure}[htbp]
\begin{center}
\includegraphics[width=.9\linewidth]{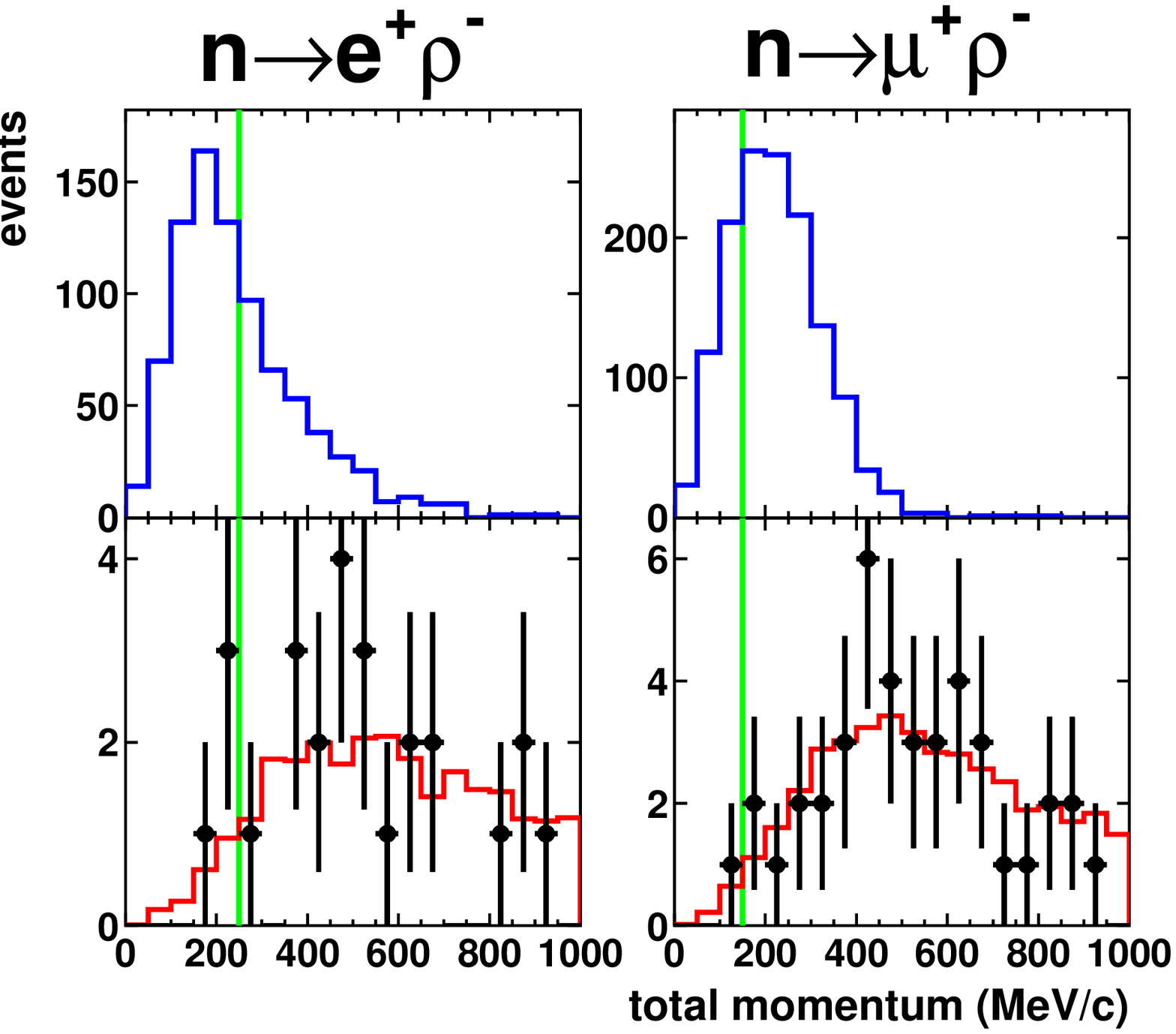}
\end{center}
\caption{ \protect \small
Total momentum for $n \rightarrow l^+ \rho^-$ MC (upper), atmospheric neutrino MC (lower, red histogram), and data (lower, black circles) for $n \rightarrow e^+ \rho^-$ (left) and $n \rightarrow \mu^+ \rho^-$ (right) searches.
All the event selections except (G6) in Section~\ref{sec:n2lrho} are applied.
The atmospheric neutrino MC is normalized to data by area.
Results from SK-I to SK-IV are combined.
The signal region corresponds to left side from the vertical green line.
Total momenta of the data candidates are 173~MeV/c, 235~MeV/c, 239~MeV/c, 246~MeV/c for $n \rightarrow e^+ \rho^-$ search and 129~MeV/c for $n \rightarrow \mu^+ \rho^-$ search.
}
\label{fig:n2lrho_ptot}
\end{figure}

%Figure~\ref{fig:n2erho_nevts}--\ref{fig:n2murho_nevts} show the signal selection efficiency and the number of expected background events at each event selection step. 
Figure~\ref{fig:n2lrho_nevts} shows the signal selection efficiency and the number of expected background events at each event selection step. 
%
%
%The total mass and total momentum distributions are shown in Figure~\ref{fig:n2erho_2d}--\ref{fig:n2murho_2d} and Figure~\ref{fig:n2erho_ptot}--\ref{fig:n2murho_ptot}.
The total mass and total momentum distributions are shown in Fig.~\ref{fig:n2lrho_2d} and Fig.~\ref{fig:n2lrho_ptot}.

There are four data candidates (one in SK-I, one in SK-II, and two in SK-III) in $n \rightarrow e^+ \rho^-$ search and one data candidate in SK-II in $n \rightarrow \mu^+ \rho^-$ search. The number of total expected background is 0.87 for $n \rightarrow e^+ \rho^-$ and 0.96 for $n \rightarrow \mu^+ \rho^-$ searches. The Poisson probability to observe four or more events is 1.2\% for $n \rightarrow e^+ \rho^-$ search. This is the lowest probability among all the nucleon decay searches and more details of these four events are described in Section~\ref{sec:results}. 
%The event displays are shown in Figure~\ref{fig:n2erho_ev1}--\ref{fig:n2murho_ev1}.

The one data candidate in SK-I in the $n \rightarrow e^+ \rho^-$
search in this analysis was also selected in the previous analysis.
There is one SK-II data candidate in both the $n \rightarrow e^+
\rho^-$ and $n \rightarrow \mu^+ \rho^-$ searches in this
analysis. Both events were not selected in the previous analysis,
because their total momentum and mass were near the cut thresholds but
outside of the signal box.

%\clearpage

%\subsection{Search results in the SK data}\label{sec:results}
\section{Search results in the SK data}\label{sec:results}

The signal selection efficiency, the number of the expected atmospheric neutrino backgrounds, and the number of data candidate events for each SK period are summarized in Table~\ref{tab:summaryall}.
\begin{table*}[htbp]
\begin{center}
\begin{tabular}{lcccccccccccc}
\hline \hline
& \multicolumn{4}{c}{Efficiency} 
& \multicolumn{4}{c}{Background} 
& \multicolumn{4}{c}{Candidate}
\\
& \multicolumn{4}{c}{(\%)} 
& \multicolumn{4}{c}{(events)} 
& \multicolumn{4}{c}{(events)}
\\
Modes & I & II & III & IV & I & II & III & IV & I & II & III & IV\\
\hline
$p\rightarrow e^+\eta$ & & & & & & & & & & & & \\
{\scriptsize (2$\gamma$, upper)} & 11.0 & 10.9 & 10.7 & 9.8 & 0.17$\pm$0.04(1.8) & 0.10$\pm$0.02(2.0) & 0.05$\pm$0.01(1.5) & 0.13$\pm$0.04(0.92) & 0 & 0 & 0 & 0 \\
{\scriptsize (2$\gamma$, lower)} & 7.9 & 6.7 & 8.2 & 7.5 & 0.01$\pm$0.01(0.09) & 0.01$\pm$0.01(0.18) & 0.003$\pm$0.003(0.09) & 0.01$\pm$0.01(0.09) & 0 & 0 & 0 & 0 \\
{\scriptsize (3$\pi^0$)} & 8.0 & 8.2 & 7.6 & 7.7 & 0.15$\pm$0.03(1.6) & 0.06$\pm$0.02(1.1) & 0.06$\pm$0.01(2.0) & 0.03$\pm$0.02(0.19) & 0 & 0 & 0 & 0 \\
$p\rightarrow \mu^+\eta$ & & & & & & & & & & & & \\
{\scriptsize (2$\gamma$, upper)} & 7.3 & 6.5 & 7.2 & 8.4 & 0.05$\pm$0.02(0.54) & 0.02$\pm$0.01(0.31) & 0.01$\pm$0.01(0.28) & 0.03$\pm$0.01(0.17) & 0 & 0 & 0 & 0 \\
{\scriptsize (2$\gamma$, lower)} & 5.8 & 5.6 & 6.0 & 7.0 & 0+0.006(0) & 0+0.004(0) & 0+0.003(0) & 0+0.008(0) & 0 & 0 & 0 & 0 \\
{\scriptsize (3$\pi^0$)} & 6.9 & 6.2 & 6.9 & 7.9 & 0.34$\pm$0.05(3.7) & 0.13$\pm$0.02(2.7) & 0.12$\pm$0.02(3.7) & 0.16$\pm$0.04(1.1) & 0 & 1 & 0 & 1 \\
\hline
$p\rightarrow e^+\rho^0$ & 3.8 & 3.3 & 3.6 & 3.8 & 0.29$\pm$0.05(3.2) & 0.13$\pm$0.02(2.6) & 0.05$\pm$0.01(1.4) & 0.17$\pm$0.04(1.2) & 0 & 0 & 0 & 2 \\
$p\rightarrow \mu^+\rho^0$ & 1.9 & 1.3 & 2.2 & 1.9 & 0.41$\pm$0.05(4.4) & 0.21$\pm$0.03(4.4) & 0.13$\pm$0.02(4.0) & 0.55$\pm$0.07(3.8) & 1 & 0 & 0 & 0 \\
\hline
$p\rightarrow e^+\omega$ & & & & & & & & & & & & \\
{\scriptsize ($\pi^0\gamma$)} & 2.3 & 2.5 & 2.3 & 2.1 & 0.16$\pm$0.04(1.7) & 0.08$\pm$0.02(1.6) & 0.06$\pm$0.01(1.8) & 0.05$\pm$0.03(0.36) & 0 &0 & 0 & 0 \\
{\scriptsize ($\pi^+\pi^-\pi^0$)} & 2.7 & 2.2 & 2.6 & 2.7 & 0.44$\pm$0.06(4.8) & 0.17$\pm$0.03(3.4) & 0.12$\pm$0.02(3.9) & 0.27$\pm$0.06(1.8) & 1 &0 & 0 & 0 \\
$p\rightarrow \mu^+\omega$ & & & & & & & & & & & & \\
{\scriptsize ($\pi^0\gamma$)} & 2.6 & 3.0 & 3.1 & 3.3 & 0.18$\pm$0.04(1.9) & 0.10$\pm$0.02(1.1) & 0.08$\pm$0.01(2.4) & 0.07$\pm$0.03(0.71) & 0 &0 & 0 & 0 \\
{\scriptsize ($\pi^+\pi^-\pi^0$)} & 3.1 & 2.6 & 3.2 & 4.6 & 0.19$\pm$0.03(2.0) & 0.10$\pm$0.02(1.1) & 0.08$\pm$0.01(2.5) & 0.29$\pm$0.05(2.0) & 0 &0 & 0 & 0 \\
\hline
$n\rightarrow e^+\pi^-$ & 12.7 & 12.2 & 13.5 & 12.6 & 0.17$\pm$0.04(1.9) & 0.05$\pm$0.01(1.1) & 0.07$\pm$0.01(2.0) & 0.12$\pm$0.04(0.83) & 0 &0 & 0 & 0 \\ 
$n\rightarrow \mu^+\pi^-$ & 11.3 & 10.7 & 11.5 & 13.4 & 0.29$\pm$0.04(3.1) & 0.15$\pm$0.02(3.0) & 0.09$\pm$0.01(2.9) & 0.24$\pm$0.05(1.7) & 0 &0 & 1 & 0 \\
\hline
$n\rightarrow e^+\rho^-$ & 1.4 & 1.1 & 1.4 & 1.5 & 0.36$\pm$0.05(3.9) & 0.13$\pm$0.02(2.7) & 0.14$\pm$0.02(4.4) & 0.24$\pm$0.06(1.6) & 1 & 1 & 2 & 0 \\
$n\rightarrow \mu^+\rho^-$ & 1.0 & 1.0 & 1.1 & 1.2 & 0.34$\pm$0.04(3.7) & 0.14$\pm$0.02(2.8) & 0.14$\pm$0.02(4.3) & 0.34$\pm$0.06(2.4) & 0 & 1 & 0 & 0 \\
\hline \hline
\end{tabular}
\caption{
Summary of the signal efficiencies, the number of expected backgrounds, and the number of data candidates for 91.5, 49.1, 31.8, and 143.8 kiloton$\cdot$year exposure during SK-I, SK-II, SK-III, and SK-IV.
For $p \rightarrow l^+ \eta$, $\eta \rightarrow 2\gamma$ search, ``upper'' and ``lower'' stand for 100$<P_\textrm{tot}<$250~MeV/c and $P_\textrm{tot}<$100~MeV/c, respectively.
Errors on the backgrounds are statistical from finite MC statistics (about 500~years for each SK period).
The number of expected backgrounds scaled to 1 megaton$\cdot$year are shown in parentheses.
}
\label{tab:summaryall}
\end{center}
\end{table*}
For all the nucleon decay searches, data and the atmospheric neutrino
background MC agree with each other in both the event rates along the
event selections and the relevant selection parameter distribution
shapes shown in Section~\ref{sec:selections}. There is no significant
excess of data above the background expectation.

% detector dependence of efficiency and #BKG
Thanks to the improvements of the 2008 electronics and DAQ upgrade,
the Michel electron tagging efficiency is higher and hence signal
efficiency is improved in SK-IV, especially for nucleon decay modes
including $\mu^+$ in the final state. Likwise, the addition of neutron
tagging has reduced the atmospheric neutrino background rate in SK-IV
compared to SK-I-II-III while maintaining similar or higher signal
efficiency.

The classification of background atmospheric neutrino interactions is
shown in Table~\ref{tab:breakdown}. They are consistent with the
previous analysis (Table~V in \cite{Nishino:2012bnw}) within the
statistical errors.

% note Nishino did not show the breakdowns for all the nucleon decay modes.

\begin{table}[htbp]
\begin{center}
\begin{tabular}{lccccc}
\hline \hline
Mode & $p \rightarrow l^+ \eta$ & $p \rightarrow l^+ \rho$ & $p \rightarrow l^+ \omega$ & $n \rightarrow l^+ \pi$ & $n \rightarrow l^+ \rho$\\
\hline
CCQE           &  4 &  9 &  1 & 22 &  4 \\ 
CC 1-$\pi$     & 27 & 59 & 27 & 53 & 45 \\
CC multi-$\pi$ & 21 & 10 & 32 & 11 & 13 \\
CC others      & 23 &  1 &  8 &  3 &  3 \\
NC             & 25 & 21 & 32 & 12 & 35 \\
\hline \hline
\end{tabular}
\caption{The breakdown (percentage contribution) of the neutrino interaction modes of the background events~\cite{Hayato:2002sd, Mitsuka:2007zz, Mitsuka:2008zza}.
The breakdowns are calculated by adding background events of the modes decaying into each meson from SK-I to SK-IV in total.
}
\label{tab:breakdown}
\end{center}
\end{table}

% data candidates
%All the event displays ({\it supplemental material}) are inspected.
%SOME EVENT DISPLAYS WILL BE CHOSEN.
%ALL EVENT DISPLAYS WILL BE SUBMITTED TO SUPPLEMENTAL MATERIAL.

There are twelve data candidates in total (all the event displays in
\cite{temp:suppl}). Back-to-back Cherenkov rings are visible and
 events are reconstructed as expected. All of the candidate events
are near the total mass and momentum cut thresholds. For the SK-I and
SK-II data samples, which have been reanalyzed for this paper, event
migrations between the previous and current analysis can be explained
by slight differences of the reconstructed parameters due to updates
of the reconstruction algorithms as well as the detector calibrations.

% data/BKG agreement -> Poisson prob.
The number of expected backgrounds and data candidates as well as
their Poisson probabilities are summarized in
Table~\ref{tab:summaryall2}. The number of detected events is
consistent with the expected atmospheric neutrino background for all
the searches.

\begin{table*}[htbp]
\begin{center}
\begin{tabular}{lcccc}
\hline \hline
Modes & Background & Candidate & Probability & Lifetime Limit \\
      & (events)   & (events)  &  (\%)       & ($\times$10$^{33}$years) at 90\% CL\\
\hline
% summed #BKG for SK-I+II from N's paper
%$p\rightarrow e^+\eta$     & 0.78(0.44) & 0(0) & -  & 12 & 10.(4.2)\\
%$p\rightarrow \mu^+\eta$   & 0.85(0.49) & 2(2) & 20.9 & 12 & 4.7(1.3)\\
%$p\rightarrow e^+\rho^0$     & 0.64(0.35) & 2(0) & 13.5 & 4  & 0.72(0.71)\\
%$p\rightarrow \mu^+\rho^0$   & 1.30(0.42) & 1(1) & 72.7 & 4  & 0.57(0.16)\\
%$p\rightarrow e^+\omega$   & 1.35(0.53) & 1(1) & 74.1 & 4  & 1.6(0.32)\\
%$p\rightarrow \mu^+\omega$ & 1.09(0.48) & 0(0) & -  & 4  & 2.8(0.78)\\
%$n\rightarrow e^+\pi^-$      & 0.41(0.27) & 0(0) & -  & 4  & 5.3(2.0)\\
%$n\rightarrow \mu^+\pi^-$    & 0.77(0.43) & 1(1) & 53.7 & 4  & 3.5(1.0)\\
%$n\rightarrow e^+\rho^-$     & 0.87(0.38) & 4(1) & 1.2  & 4  & 0.03(0.07)\\
%$n\rightarrow \mu^+\rho^-$   & 0.96(0.29) & 1(0) & 61.7 & 4  & 0.06(0.04)\\
$p\rightarrow e^+\eta$     & 0.78$\pm$0.30 & 0 & -  & 10.\\
$p\rightarrow \mu^+\eta$   & 0.85$\pm$0.23 & 2 & 20.9 & 4.7\\
$p\rightarrow e^+\rho^0$     & 0.64$\pm$0.17 & 2 & 13.5 & 0.72\\
$p\rightarrow \mu^+\rho^0$   & 1.30$\pm$0.33 & 1 & 72.7 & 0.57\\
$p\rightarrow e^+\omega$   & 1.35$\pm$0.43 & 1 & 74.1 & 1.6\\
$p\rightarrow \mu^+\omega$ & 1.09$\pm$0.52 & 0 & -  & 2.8\\
$n\rightarrow e^+\pi^-$      & 0.41$\pm$0.13 & 0 & -  & 5.3\\
$n\rightarrow \mu^+\pi^-$    & 0.77$\pm$0.20 & 1 & 53.7 & 3.5\\
$n\rightarrow e^+\rho^-$     & 0.87$\pm$0.26 & 4 & 1.2  & 0.03\\
$n\rightarrow \mu^+\rho^-$   & 0.96$\pm$0.28 & 1 & 61.7 & 0.06\\
\hline
%total                        & 8.6(4.1) & 12(6)   & 15.7 & - & -\\
total                        & 8.6 & 12 & 15.7 & -\\
\hline \hline
\end{tabular}
\caption{ Summary of the nucleon decay searches.  The number of events
  are summed from SK-I to SK-IV and for all the meson decay modes for
  each nucleon decay search.  The events in the total signal box
  are shown for $p \rightarrow l^+
  \eta$ search.  
The number of the expected atmospheric backgrounds with the systematic errors (see Tab.~\ref{tab:bkgerr} for more details) are shown in the second column.
  The Poisson probability to observe events greater than or equal to
  the number of data candidates,
without considering the systematic uncertainty in the background,
is shown in the fourth column.  
%The number of measurements used in the lifetime limit calculation (Sec.~\ref{sec:limit}) are shown in the fifth column.
%Numbers in parentheses are from the previous analysis in SK-I and SK-II.
}
\label{tab:summaryall2}
\end{center}
\end{table*}

The lowest probability (1.2\%) is seen in the $n \rightarrow e^+
\rho^-$ search. The number of events for each event selection step
(Fig.~\ref{fig:n2lrho_nevts}) is in agreement with expectation leading
up to the application of the momentum and mass signal region. In
total, there are four data candidates. Their total invariant mass
(MeV/c$^2$) and total momentum (MeV/c) are (884.7, 245.5), (821.5,
239.1), (897.7, 235.4), and (807.8, 173.2), respectively. Every
candidate event is near the threshold of the selection window of the
total mass and the momentum cut. The total invariant mass and momentum
plots (Figs.~\ref{fig:n2lrho_2d} and \ref{fig:n2lrho_ptot}) show that
the data distributions agree with the expected atmospheric neutrino
background distributions. Overall, we have concluded that the
observed data candidates are due to atmospheric neutrino backgrounds.

% comparison of eff/#BKG/candidates in SK-I/II with Nishino Although
%the total signal box $P_{tot}<$205~MeV/c is not used in this analysis
%for $p \rightarrow l^+ \eta$, $\eta \rightarrow 2\gamma$ search,
%the signal efficiencies, the number of expected background events,
%and the number of data candidates are shown in
%Table~\ref{tab:summary12} to compare with the previous study.

The estimated signal efficiencies for $n \rightarrow l^+ \pi^-$ with
the charged pion in the final state are lower in this analysis than
the previous analysis. For example, the signal efficiency for the $n
\rightarrow e^+ \pi^-$ search in SK-I decreased from 19.4\% to
12.7\%. This is due to the improved pion interaction model which
better matches external data. As shown in Fig.~\ref{fig:pimxsec}, the
cross sections for all the interactions are increased around
500~MeV/c. The larger cross sections reduce the number of the visible
Cherenkov rings, with reduction of the estimated signal efficiency
mainly coming from the corresponding event selection.

\subsection{Systematic errors}\label{sec:syserr}

The total systematic errors on the signal efficiency and the number of
expected background events are summarized in
Table~\ref{tab:sigerr}--\ref{tab:bkgerr}.
The systematic error sources consist of uncertainties of physics
models in the nucleon decay MC, the atmospheric neutrino background
MC, and event reconstruction performance.  The systematic error
estimation methods are the same as the previous analysis, with several
improvements that allow to significantly reduce the errors.

%Table~\ref{tab:sigerr12} shows the systematic errors on the signal efficiencies for each systematic error source in SK-I and SK-II.
% To directly compare with previous paper, I take an average of SK-I(91.5ktyr) and SK-II(49.1ktyr) for detector performances. The weights SK-I: 91.5/(91.5+49.1)=0.65, SK-II: 1-0.65=0.35. Errors from mine_ndk_limit20160728.pdf for each SK period.
%%% 20170114 calc average for sk1-4 weighted by livetime
Table~\ref{tab:sigerr} shows the systematic errors on the signal efficiencies for each systematic error source.
\begin{table*}[htbp]
\begin{center}
\begin{tabular}{lcccccc}
\hline \hline
Modes & Meson & Hadron propagation & N-N correlated & Fermi & Detector & \\
 & nuclear effect & in water & decay & momentum & performances & Total \\
\hline
%$p\rightarrow e^+\eta$ & & & & & &\\
%{\scriptsize (2$\gamma$, total)} & 19(20) & - & 6(7) & 0.4(13) & 3(5) & 20(25) \\
%{\scriptsize (2$\gamma$, upper)} & 25 & - & 8 & 9 & 3 & 28 \\
%{\scriptsize (2$\gamma$, lower)} & 10 & - & 3 & 13 & 3 & 17 \\
$p\rightarrow e^+\eta$ {\scriptsize (2$\gamma$, upper)} & 26 & - & 8 & 9 & 2 & 29 \\
$p\rightarrow e^+\eta$ {\scriptsize (2$\gamma$, lower)} & 9 & - & 3 & 13 & 2 & 16 \\
%
%{\scriptsize (3$\pi^0$)} & 14(15) & - & 4(5) & 15(26) & 3(9) & 21(32) \\
$p\rightarrow e^+\eta$ {\scriptsize (3$\pi^0$)} & 12 & - & 4 & 15 & 4 & 20 \\
%
%$p\rightarrow \mu^+\eta$ & & & & & &\\
%{\scriptsize (2$\gamma$, total)} & 14(18) & - & 6(7) & 2(14) & 3(4) & 16(24) \\
%{\scriptsize (2$\gamma$, upper)} & 26 & - & 9 & 13 & 3 & 31 \\
%{\scriptsize (2$\gamma$, lower)} & 7 & - & 3 & 12 & 3 & 14 \\
$p\rightarrow \mu^+\eta$ {\scriptsize (2$\gamma$, upper)} & 27 & - & 9 & 10 & 3 & 30 \\
$p\rightarrow \mu^+\eta$ {\scriptsize (2$\gamma$, lower)} & 11 & - & 3 & 12 & 3 & 17 \\
%{\scriptsize (3$\pi^0$)} & 16(20) & - & 6(7) & 3(14) & 4(10) & 18(28) \\
$p\rightarrow \mu^+\eta$ {\scriptsize (3$\pi^0$)} & 17 & - & 6 & 2 & 5 & 19 \\
%\hline
%$p\rightarrow e^+\rho^0$ & 9(8) & 13(17) & 0.6(2) & 4(10) & 5(18) & 17(28) \\
%$p\rightarrow \mu^+\rho^0$ & 10(9) & 9(24) & 2(2) & 1(6) & 11(11) & 18(29) \\
$p\rightarrow e^+\rho^0$ & 9 & 13 & 1 & 4 & 6 & 17 \\
$p\rightarrow \mu^+\rho^0$ & 10 & 9 & 1 & 2 & 10 & 17 \\
%\hline
%$p\rightarrow e^+\omega$ & & & & & &\\
%{\scriptsize ($\pi^0\gamma$)} & 14(21) & - & 3(5) & 13(24) & 4(9) & 20(33) \\
$p\rightarrow e^+\omega$ {\scriptsize ($\pi^0\gamma$)} & 14 & - & 3 & 14 & 4 & 20 \\
%{\scriptsize ($\pi^+\pi^-\pi^0$)} & 21(19) & 6(13) & 3(5) & 0.7(12) & 6(20) & 22(34) \\
$p\rightarrow e^+\omega$ {\scriptsize ($\pi^+\pi^-\pi^0$)} & 14 & 6 & 3 & 0.4 & 7 & 17 \\
%
%$p\rightarrow \mu^+\omega$ & & & & & &\\
%{\scriptsize ($\pi^0\gamma$)} & 19(23) & - & 1(6) & 2(13) & 4(7) & 19(28) \\
$p\rightarrow \mu^+\omega$ {\scriptsize ($\pi^0\gamma$)} & 14 & - & 1 & 2 & 3 & 14 \\
%{\scriptsize ($\pi^+\pi^-\pi^0$)} & 26(19) & 5(15) & 2(5) & 1(2) & 5(16) & 27(29) \\
$p\rightarrow \mu^+\omega$ {\scriptsize ($\pi^+\pi^-\pi^0$)} & 24 & 5 & 2 & 1 & 4 & 25 \\
%\hline
%$n\rightarrow e^+\pi^-$ & 13(20) & 7(9) & 11(11) & 0.8(12) & 4(12) & 19(30) \\
$n\rightarrow e^+\pi^-$ & 13 & 7 & 11 & 1 & 4 & 19 \\
%$n\rightarrow \mu^+\pi^-$ & 15(24) & 7(6) & 10(11) & 3(7) & 4(17) & 20(33) \\
$n\rightarrow \mu^+\pi^-$ & 15 & 7 & 10 & 3 & 4 & 20 \\
%\hline
%$n\rightarrow e^+\rho^-$ & 60(+51, -23) & 3(9) & 7(11) & 9(15) & 6(19) & 61(+59, -37) \\
$n\rightarrow e^+\rho^-$ & 60 & 3 & 9 & 5 & 5 & 61 \\
%$n\rightarrow \mu^+\rho^-$ & 48(+51, -25) & 5(14) & 0.6(10) & 15(27) & 7(23) & 51(+65, -47) \\
$n\rightarrow \mu^+\rho^-$ & 48 & 5 & 3 & 12 & 8 & 50 \\
\hline \hline
\end{tabular}
\caption{
%Summary of systematic errors on signal efficiencies for each error source in SK-I and SK-II.
%The errors of SK-I and SK-II are averaged by the livetime.
%For $p \rightarrow l^+ \eta$, $\eta \rightarrow 2\gamma$ search, ``total'', ``upper'', and ``lower'' stand for $P_{tot}<$250~MeV/c, 100$<P_{tot}<$250~MeV/c, and $P_{tot}<$100~MeV/c, respectively.
Summary of systematic errors (percentage contribution) on signal efficiencies for each error source.
The errors from SK-I to SK-IV are averaged by the livetime.
For $p \rightarrow l^+ \eta$, $\eta \rightarrow 2\gamma$ search, ``upper'' and ``lower'' stand for 100$<P_\textrm{tot}<$250~MeV/c and $P_\textrm{tot}<$100~MeV/c, respectively.
%Numbers in parentheses are from the previous analysis.
}
\label{tab:sigerr}
\end{center}
\end{table*}
%
%Common errors for SK-I,II,III for physics models.
%
The dominant systematic errors come from uncertainties of the meson interactions in the oxygen nucleus (meson nuclear effect).
The meson nuclear effect errors are smaller in the lower signal box for $p \rightarrow l^+ \eta$ and $\eta \rightarrow 2\gamma$ searches thanks to the large fraction of the free proton decays.
%

% ----- Hide.T added (2016/10/26) ------------
%NEUT neutrino interaction simulation~\cite{{Hayato:2002sd},{Mitsuka:2007zz},{Mitsuka:2008zza}} is adopted in this analysis in order to simulate the pion interactions in oxygen nuclear medium and the pion interactions in the detector medium.
%NEUT is adopted in this analysis in order to simulate the pion interactions in oxygen nuclei and the pion interactions in water.
%The pion interactions include inelastic scattering, charge exchange and absorption, as well as particle production for high energy pions.
% ----- Hide.T updated 'pion scattering model' description (2017/02/06)
% (according to Hayato-san's comments)
%The pion interaction models in NEUT is a semi-classical cascade model.
%The interaction probability (mean free paths) for low momentum pions are calculated with the Delta-hole model by Oset {\it et al}~\cite{Salcedo:1987md}.
%The pion interaction probability for high momentum pions are extracted from the $\pi^{\pm}$-nucleus scattering experimental data.
%All the pion interaction probabilities and cross section parameters have been tuned with various $\pi^{\pm}$-nucleus scattering data including C, O, Al, Fe~\cite{Miura:2016lpi0} since the previous paper~\cite{Nishino:2012bnw}.
%The pion interaction models in NEUT, which is based on the Bertini intra-nuclear hadronic cascade model~\cite{Bertini:1970zs}, has been tuned with various $\pi^{\pm}$-nucleus scattering data including C, O, Al, Fe~\cite{Miura:2016lpi0} since the previous paper~\cite{Nishino:2012bnw}.
In this analysis, the systematic uncertainties of pion interactions are evaluated by varying the tuning parameters of the interaction models within 1$\sigma$ uncertainties of $\pi^{\pm}$-nucleus experimental data.
%This method provides a more reliable evaluation of systematic uncertainties of the pion interactions and reduces their systematic uncertainties compared to the previous paper~\cite{Nishino:2012bnw}, especially for higher momentum ($\geq500$~MeV/c$^2$) pions, that causes relatively smaller systematic uncertainties of the pion interactions in atmospheric neutrino background events than the previous analysis.
%
This method provides a more reliable evaluation of systematic uncertainties of the pion interactions compared to the previous paper~\cite{Nishino:2012bnw}, especially for higher momentum ($\geq500$~MeV/c) pions. %This way leads relatively smaller systematic uncertainties of the pion interactions in atmospheric neutrino background events than the previous analysis. 
This method leads to relatively smaller systematic uncertainties of the pion interactions in signal and atmospheric neutrino background events than the previous analysis. 
% --------------------------------------------

%
%The uncertainty of the Fermi momentum was conservatively estimated in the previous analysis. We compared a spectral fit based on data scattering model~\cite{Nakamura:1976mb} vs. Fermi gas model but the Fermi momentum distribution shape was artificially scaled by 20\% in the previous analysis.
%
In the previous analysis~\cite{Nishino:2012bnw} the uncertainty in
Fermi momentum distribution was estimated by artificially re-scaling
the distribution by 20\%. While being simple, this is a rather crude
procedure. In this work, we improve and provide a more precise
uncertainty estimate by directly comparing the momentum distributions,
simulated using a spectral function fit to the $^{12}$C electron
scattering data~\cite{Nakamura:1976mb}, to the simulated atmospheric
background momentum distribution from NEUT, which uses a relativistic
Fermi gas model. The difference between the two is taken as a
systematic error.

%
%Uncertainties of the event reconstructions used in the event selections are taken into account as the detector performance errors. The error sources are vertex, the number of Cherenkov rings, particle identification (e/$\mu$ separation), the number of Michel electrons, and the energy scale.
%The additional conservative error estimation, artificially shifting the vertex by 30~cm, was employed in the previous analysis but not used in this analysis. This is the main reason of reduction of the detector performance errors.

Uncertainties associated with event reconstruction are implemented as
detector performance errors. The error sources include: vertex
position, number of Cherenkov rings, particle identification (e/$\mu$
separation), the number of Michel electrons, and the energy scale. In
the previous study, the error estimate was overly conservative and
included an additional contribution from the effects of artificially
shifting the vertex by 30~cm. We do not adopt such contribution in
this work.

%which is the main reason for the reduction of the detector performance error.

%Common errors for total/upper/lower box for detector performances.
%

%Table~\ref{tab:bkgerr12} shows the systematic errors on the number of expected background events for each systematic error source in SK-I and SK-II.
% To directly compare with previous paper, I take an average of SK-I(91.5ktyr) and SK-II(49.1ktyr) for detector performances. The weights SK-I: 91.5/(91.5+49.1)=0.65, SK-II: 1-0.65=0.35. Errors from mine_ndk_limit20160728.pdf for each SK period.

Table~\ref{tab:bkgerr} shows the systematic errors on the number of
expected background events for each systematic error source. After the
event selection, the raw number of the remaining expected backgrounds
is either zero or one in the lower signal box for $p \rightarrow l^+
\eta$, $\eta \rightarrow 2\gamma$ searches. Therefore, the
systematic errors are estimated in the total box ($P_\textrm{tot} <$
250~MeV/c) and assumed to be same both in the upper and lower boxes.
With the improved error estimation method, the systematic errors on
the most dominant systematic error source in the previous analysis,
hadron propagation in water, are reduced by factor 3-5 in this
analysis.

\begin{table*}[htbp]
\begin{center}
\begin{tabular}{lcccccc}
\hline \hline
Modes & Neutrino & Neutrino cross & Pion nuclear & Hadron propagation & Detector & \\
 & flux & section & effect & in water & performances & Total \\
\hline
%$p\rightarrow e^+\eta$ & & & & & &\\
%{\scriptsize (2$\gamma$, total)} & 8(8) & 14(11) & 12(5) & 5(36) & 20(26) & 29(47) \\
%{\scriptsize (2$\gamma$, upper)} & 8 & 14 & 12 & 5 & 20 & 29 \\
%{\scriptsize (2$\gamma$, lower)} & 8 & 14 & 12 & 5 & 20 & 29 \\
$p\rightarrow e^+\eta$ {\scriptsize (2$\gamma$, upper)} & 8 & 15 & 12 & 5 & 21 & 30 \\
$p\rightarrow e^+\eta$ {\scriptsize (2$\gamma$, lower)} & 8 & 15 & 12 & 5 & 21 & 30 \\
%
%{\scriptsize (3$\pi^0$)} & 8(8) & 15(15) & 34(18) & 17(67) & 30(13) & 51(76) \\
$p\rightarrow e^+\eta$ {\scriptsize (3$\pi^0$)} & 8 & 12 & 34 & 17 & 30 & 51 \\
%
%$p\rightarrow \mu^+\eta$ & & & & & &\\
%{\scriptsize (2$\gamma$, total)} & 8(8) & 11(14) & 14(5) & 8(36) & 27(28) & 35(49) \\
%{\scriptsize (2$\gamma$, upper)} & 8 & 11 & 14 & 8 & 27 & 35 \\
%{\scriptsize (2$\gamma$, lower)} & 8 & 11 & 14 & 8 & 27 & 35 \\
$p\rightarrow \mu^+\eta$ {\scriptsize (2$\gamma$, upper)} & 9 & 15 & 14 & 8 & 28 & 37 \\
$p\rightarrow \mu^+\eta$ {\scriptsize (2$\gamma$, lower)} & 9 & 15 & 14 & 8 & 28 & 37 \\
%
%{\scriptsize (3$\pi^0$)} & 8(8) & 11(11) & 9(18) & 13(67) & 11(20) & 23(73) \\
$p\rightarrow \mu^+\eta$ {\scriptsize (3$\pi^0$)} & 8 & 12 & 9 & 13 & 12 & 25 \\
%\hline
%$p\rightarrow e^+\rho^0$ & 5(6) & 15(13) & 10(14) & 7(33) & 13(33) & 24(51) \\
%$p\rightarrow \mu^+\rho^0$ & 7(8) & 16(15) & 5(14) & 11(33) & 12(23) & 25(46) \\
$p\rightarrow e^+\rho^0$ & 4 & 18 & 10 & 7 & 15 & 27 \\
$p\rightarrow \mu^+\rho^0$ & 7 & 16 & 5 & 11 & 14 & 25 \\
%\hline
%$p\rightarrow e^+\omega$ & & & & & &\\
%{\scriptsize ($\pi^0\gamma$)} & 8(8) & 16(14) & 21(13) & 5(41) & 31(37) & 42(59) \\
%{\scriptsize ($\pi^+\pi^-\pi^0$)} & 7(7) & 15(14) & 7(8) & 14(53) & 15(28) & 27(63) \\
%
$p\rightarrow e^+\omega$ {\scriptsize ($\pi^0\gamma$)} & 8 & 15 & 21 & 5 & 32 & 42 \\
$p\rightarrow e^+\omega$ {\scriptsize ($\pi^+\pi^-\pi^0$)} & 6 & 15 & 7 & 14 & 17 & 28 \\
%
%$p\rightarrow \mu^+\omega$ & & & & & &\\
%{\scriptsize ($\pi^0\gamma$)} & 8(8) & 12(10) & 18(13) & 8(41) & 27(28) & 36(53) \\
%{\scriptsize ($\pi^+\pi^-\pi^0$)} & 8(7) & 10(11) & 22(8) & 46(53) & 14(29) & 55(63) \\
$p\rightarrow \mu^+\omega$ {\scriptsize ($\pi^0\gamma$)} & 8 & 13 & 18 & 8 & 28 & 37 \\
$p\rightarrow \mu^+\omega$ {\scriptsize ($\pi^+\pi^-\pi^0$)} & 8 & 10 & 22 & 46 & 16 & 55 \\
%\hline
%$n\rightarrow e^+\pi^-$ & 7(8) & 14(15) & 6(8) & 10(36) & 24(46) & 31(61) \\
%$n\rightarrow \mu^+\pi^-$ & 9(8) & 17(16) & 3(8) & 6(36) & 16(36) & 25(55) \\
$n\rightarrow e^+\pi^-$ & 7 & 17 & 6 & 10 & 24 & 32 \\
$n\rightarrow \mu^+\pi^-$ & 8 & 16 & 3 & 6 & 17 & 26 \\
%\hline
%$n\rightarrow e^+\rho^-$ & 7(8) & 15(14) & 15(12) & 16(18) & 9(54) & 29(60) \\
%$n\rightarrow \mu^+\rho^-$ & 6(6) & 19(16) & 15(12) & 4(18) & 13(27) & 28(39) \\
$n\rightarrow e^+\rho^-$ & 7 & 16 & 15 & 16 & 11 & 30 \\
$n\rightarrow \mu^+\rho^-$ & 5 & 20 & 15 & 4 & 14 & 29 \\
\hline \hline
\end{tabular}
\caption{
%Summary of systematic errors on the number of expected background for each error source in SK-I and SK-II.
%The errors of SK-I and SK-II are averaged by the livetime.
%For $p \rightarrow l^+ \eta$, $\eta \rightarrow 2\gamma$ searches, ``total'', ``upper'', and ``lower'' stand for $P_{tot}<$250~MeV/c, 100$<P_{tot}<$250~MeV/c, and $P_{tot}<$100~MeV/c, respectively.
Summary of systematic errors (percentage contribution) on the number of expected background for each error source.
The errors from SK-I to SK-IV are averaged by the livetime.
For $p \rightarrow l^+ \eta$, $\eta \rightarrow 2\gamma$ searches, ``upper'' and ``lower'' stand for 100$<P_\textrm{tot}<$250~MeV/c and $P_\textrm{tot}<$100~MeV/c, respectively.
%Numbers in parentheses are from the previous analysis.
%In the previous analysis, an average of the nucleon decay searches with the $e^+$ and $\mu^+$ modes is used for pion nuclear effect and Hadron propagation in water.
}
\label{tab:bkgerr}
\end{center}
\end{table*}
%
%Common errors for SK-I,II(,III) for physics models.
%Common errors for total/upper/lower box due to few remaining events
%Detector errors are estimated for each SK detector but average is also taken
%

\subsection{Lifetime limit}\label{sec:limit}

% note: difference from N: I did not add two meson decay modes like eta->2gamma and eta->3pi0
%

All of the observed events are consistent with the expected
backgrounds and nucleon lifetime limits are calculated as in
\cite{Nishino:2012bnw,Miura:2016lpi0} using the Bayes method to
incorporate systematic uncertainty. The limits are calculated by
combining different search methods as well as different SK periods for
each nucleon decay search. 
%The number of measurements for each nucleon decay search is summarized in Table~\ref{tab:summaryall2}.  
All of the nucleon decay searches except $p \rightarrow l^+ \eta$ have four
measurements corresponding to four SK periods. As in the previous
analysis, the signal efficiencies and the backgrounds are added for $p
\rightarrow l^+ \omega$ mode with two different meson decay modes
$\omega \rightarrow \pi^0 \gamma$ and $\omega \rightarrow \pi^+ \pi^-
\pi^0$.  There are twelve measurements for $p \rightarrow l^+ \eta$
searches corresponding to three methods per each of the four SK
periods, where the three methods come from upper and lower signal
boxes for $\eta \rightarrow 2\gamma$ search and one signal box for
$\eta \rightarrow 3\pi^0$ search.

The conditional probability distribution for each nucleon decay rate
is expressed as:
\begin{eqnarray}
P(\Gamma|n_i) &=& \iiint \frac{e^{-(\Gamma\lambda_i\epsilon_i+b_i)}
(\Gamma\lambda_i\epsilon_i+b_i)^{n_i}}{n_i!} \times \nonumber \\ \nopagebreak
 & & ~~P(\Gamma)P(\lambda_i)P(\epsilon_i)P(b_i)
     \,d\lambda_i\,d\epsilon_i\,db_i,
\end{eqnarray}
where $n_{i}$ is the number of data candidate events in the $i$-th
measurement, $\lambda_i$ is the true detector exposure, $\epsilon_i$
is the true selection efficiency, and $b_i$ is the true number of
background events. The decay rate prior probability distribution
$P(\Gamma)$ is 1 for $\Gamma \ge 0$ and otherwise 0. $P(\lambda_i)$,
$P(\epsilon_i)$, and $P(b_i)$ are the prior probability distributions
for detector exposure, efficiency, and background, respectively, which
are assumed to be Gaussian distributions.
% with $\sigma$ described in Table~\ref{tab:summarysyst}. 
The uncertainty on the exposure is assumed to be 1\% for all SK periods. 

The upper limit of the nucleon decay rate, $\Gamma_{\mathrm{limit,}}$
is:
\begin{eqnarray}
\mathrm{CL}=\frac{\int^{\Gamma_{\mathrm{limit}}}_{\Gamma=0}\prod^{N}_{i=1}P(\Gamma|n_i)\,d\Gamma} {\int^\infty_{\Gamma=0}\prod^{N}_{i=1}P(\Gamma|n_i)\,d\Gamma},
\end{eqnarray}
where $N$ is the number of measurements for each nucleon decay search
and CL is the confidence level taken to be 90\%. The
partial lower lifetime limit for each nucleon decay mode is given by:
\begin{eqnarray}
\tau/\mathrm{B}=\frac{1}{\Gamma_{\mathrm{limit}}}.
\end{eqnarray}

\begin{figure}[htbp]
\begin{center}
\includegraphics[width=.9\linewidth]{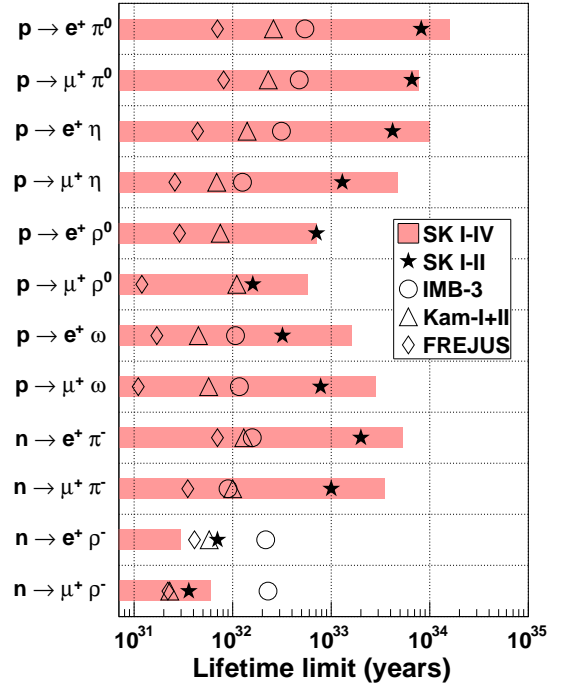}
\end{center}
\caption{ \protect \small
Explored ranges and lower limits of nucleon partial lifetime with the results of the other experiments~\cite{McGrew:1999nd, Hirata:1989kn, Berger:1990kg} and the previous SK analysis~\cite{Nishino:2012bnw}.
The $p\rightarrow l^+\pi^0$ results~\cite{Miura:2016lpi0} are also included.
}
\label{fig:limits_comp}
\end{figure}

The nucleon partial lifetime limits at 90\% confidence level are
summarized in Table~\ref{tab:summaryall2}.  The limits range from
3$\times$10$^{31}$ to 1$\times$10$^{34}$ years. Thanks to the addition
of SK-III and SK-IV data and various analysis improvements, the
lifetime limits published in this paper are typically higher than
those in the previous analysis by a factor of two to three, as shown
in Fig.~\ref{fig:limits_comp}. However, some channels show similar
or even lower limits due to more observed data candidates.

Figure~\ref{fig:limits_comp} compares the updated SK results with the
other experiments; IMB-3~\cite{McGrew:1999nd},
KAMIOKANDE-I+II~\cite{Hirata:1989kn} and FREJUS~\cite{Berger:1990kg}.
%The recent published $p\rightarrow l^+\pi^0$ results~\cite{Miura:2016lpi0} are also included. 
The updated SK results are better than those by the other experiments
by 1-2 orders of magnitude except for the $n \rightarrow l^+ \rho^0$
channel. The relatively lower performance is due to low signal
efficiencies with tight event selections and non-zero data candidates
in the SK analyses. Note that systematic errors on the signal
efficiencies were not included in the IMB-3 limit and thus the
published results were optimistic.

%\clearpage

\section{Conclusion}

The survey of nucleon decay into charged antilepton plus meson by
Super-Kamiokande based on only SK-I and SK-II data~\cite{Nishino:2012bnw} has
been updated to include improved event reconstruction algorithms,
event selections, systematic error estimations, and an additional
0.175 megaton$\cdot$years of SK-III and SK-IV data. A similar update
for the $p \rightarrow l^+ \pi^0$ search by Super-Kamiokande has also
been published recently~\cite{Miura:2016lpi0}.

Using 0.316 megaton$\cdot$years of data, no significant excess is
found above the expected atmospheric neutrino background. The number
and features of candidate events are consistent with the background
estimation by the atmospheric neutrino MC. Therefore, partial
lifetime limits are set ranging from 3$\times$10$^{31}$ to
1$\times$10$^{34}$ years at 90\% confidence level depending on the
nucleon decay mode. The lifetime limits are summarized in
Table~\ref{tab:summaryall2}.

Most of the results presented here are world-best limits and they
further constrain the relevant GUT models.

%{\bf Volodymyr: What is the impact of the results (on GUTs)?}

%{\it Acknowledgments.---}
%from Miura-san's epi0/mupi0 paper draft
%We gratefully acknowledge the cooperation of the Kamioka Mining and Smelting Company.  The Super-Kamiokande experiment has been built and operated from funding by the Japanese Ministry of Education, Culture, Sports, Science and Technology, the United States Department of Energy. 

\section*{ACKNOWLEDGMENTS}
We gratefully acknowledge the cooperation of the Kamioka Mining and Smelting Company. The Super-Kamiokande experiment has been built and operated from funding by the Japanese Ministry of Education, Culture, Sports, Science and Technology, the U.S. Department of Energy, and the U.S. National Science Foundation. Some of us have been supported by funds from the Research Foundation of Korea (BK21 and KNRC), the Korean Ministry of Science and Technology, the National Research Foundation of Korea (NRF-20110024009), the European Union H2020 RISE-GA641540-SKPLUS, the Japan Society for the Promotion of Science, the National Natural Science Foundation of China under Grants No. 11235006, the National Science and Engineering Research Council (NSERC) of Canada, %and the Scinet and Westgrid consortia of Compute Canada.
the Scinet and Westgrid consortia of Compute Canada,
and the National Science Centre of Poland (2015/17/N/ST2/04064, 2015/18/E/ST2/00758).


%\clearpage

\raggedright
\bibliography{reference}

\end{document}

%% file: authors-20170315.tex
\newcommand{\AFFicrr}{\affiliation{Kamioka Observatory, Institute for Cosmic Ray Research, University of Tokyo, Kamioka, Gifu 506-1205, Japan}}
\newcommand{\AFFkashiwa}{\affiliation{Research Center for Cosmic Neutrinos, Institute for Cosmic Ray Research, University of Tokyo, Kashiwa, Chiba 277-8582, Japan}}
\newcommand{\AFFipmu}{\affiliation{Kavli Institute for the Physics and
Mathematics of the Universe (WPI), The University of Tokyo Institutes for Advanced Study,
University of Tokyo, Kashiwa, Chiba 277-8583, Japan }}
\newcommand{\AFFmad}{\affiliation{Department of Theoretical Physics, University Autonoma Madrid, 28049 Madrid, Spain}}
\newcommand{\AFFubc}{\affiliation{Department of Physics and Astronomy, University of British Columbia, Vancouver, BC, V6T1Z4, Canada}}
\newcommand{\AFFbu}{\affiliation{Department of Physics, Boston University, Boston, MA 02215, USA}}
\newcommand{\AFFbnl}{\affiliation{Physics Department, Brookhaven National Laboratory, Upton, NY 11973, USA}}
\newcommand{\AFFuci}{\affiliation{Department of Physics and Astronomy, University of California, Irvine, Irvine, CA 92697-4575, USA }}
\newcommand{\AFFcsu}{\affiliation{Department of Physics, California State University, Dominguez Hills, Carson, CA 90747, USA}}
\newcommand{\AFFcnm}{\affiliation{Department of Physics, Chonnam National University, Kwangju 500-757, Korea}}
\newcommand{\AFFduke}{\affiliation{Department of Physics, Duke University, Durham NC 27708, USA}}
\newcommand{\AFFfukuoka}{\affiliation{Junior College, Fukuoka Institute of Technology, Fukuoka, Fukuoka 811-0295, Japan}}
\newcommand{\AFFgifu}{\affiliation{Department of Physics, Gifu University, Gifu, Gifu 501-1193, Japan}}
\newcommand{\AFFgist}{\affiliation{GIST College, Gwangju Institute of Science and Technology, Gwangju 500-712, Korea}}
\newcommand{\AFFuh}{\affiliation{Department of Physics and Astronomy, University of Hawaii, Honolulu, HI 96822, USA}}
\newcommand{\AFFkek}{\affiliation{High Energy Accelerator Research Organization (KEK), Tsukuba, Ibaraki 305-0801, Japan }}
\newcommand{\AFFkobe}{\affiliation{Department of Physics, Kobe University, Kobe, Hyogo 657-8501, Japan}}
\newcommand{\AFFkyoto}{\affiliation{Department of Physics, Kyoto University, Kyoto, Kyoto 606-8502, Japan}}
\newcommand{\AFFmiyagi}{\affiliation{Department of Physics, Miyagi University of Education, Sendai, Miyagi 980-0845, Japan}}
\newcommand{\AFFnagoya}{\affiliation{Institute for Space-Earth Enviromental Research, Nagoya University, Nagoya, Aichi 464-8602, Japan}}
\newcommand{\AFFkmi}{\affiliation{Kobayashi-Maskawa Institute for the Origin of Particles and the Universe, Nagoya University, Nagoya, Aichi 464-8602, Japan}}
\newcommand{\AFFpol}{\affiliation{National Centre For Nuclear Research, 00-681 Warsaw, Poland}}
\newcommand{\AFFsuny}{\affiliation{Department of Physics and Astronomy, State University of New York at Stony Brook, NY 11794-3800, USA}}
\newcommand{\AFFokayama}{\affiliation{Department of Physics, Okayama University, Okayama, Okayama 700-8530, Japan }}
\newcommand{\AFFosaka}{\affiliation{Department of Physics, Osaka University, Toyonaka, Osaka 560-0043, Japan}}
\newcommand{\AFFregina}{\affiliation{Department of Physics, University of Regina, 3737 Wascana Parkway, Regina, SK, S4SOA2, Canada}}
\newcommand{\AFFseoul}{\affiliation{Department of Physics, Seoul National University, Seoul 151-742, Korea}}
\newcommand{\AFFshizuokasc}{\affiliation{Department of Informatics in
Social Welfare, Shizuoka University of Welfare, Yaizu, Shizuoka, 425-8611, Japan}}
\newcommand{\AFFskk}{\affiliation{Department of Physics, Sungkyunkwan University, Suwon 440-746, Korea}}
\newcommand{\AFFtokyo}{\affiliation{The University of Tokyo, Bunkyo, Tokyo 113-0033, Japan }}
\newcommand{\AFFtodai}{\affiliation{Department of Physics, University of Tokyo, Bunkyo, Tokyo 113-0033, Japan }}
\newcommand{\AFFtoronto}{\affiliation{Department of Physics, University of Toronto, ON, M5S 1A7, Canada }}
\newcommand{\AFFtriumf}{\affiliation{TRIUMF, 4004 Wesbrook Mall, Vancouver, BC, V6T2A3, Canada }}
\newcommand{\AFFtokai}{\affiliation{Department of Physics, Tokai University, Hiratsuka, Kanagawa 259-1292, Japan}}
\newcommand{\AFFtsinghua}{\affiliation{Department of Engineering Physics, Tsinghua University, Beijing, 100084, China}}
\newcommand{\AFFuw}{\affiliation{Department of Physics, University of Washington, Seattle, WA 98195-1560, USA}}

\AFFicrr
\AFFkashiwa
\AFFmad
\AFFbu
\AFFubc
\AFFbnl
\AFFuci
\AFFcsu
\AFFcnm
\AFFduke
\AFFfukuoka
\AFFgifu
\AFFgist
\AFFuh
\AFFkek
\AFFkobe
\AFFkyoto
\AFFmiyagi
\AFFnagoya
\AFFkmi
\AFFpol
\AFFsuny
\AFFokayama
\AFFosaka
\AFFregina
\AFFseoul
\AFFshizuokasc
\AFFskk
\AFFtokai
\AFFtokyo
\AFFtodai
\AFFipmu
\AFFtoronto
\AFFtriumf
\AFFtsinghua
\AFFuw

%%%%%%%%%%%%%%%%%%%%%%%%%%%%%%%%%%%%%%%%%%%%%%%%%%%%%%%%%%%%%%%%%%%%
%ICRR
\author{K.~Abe}
\AFFicrr
\AFFipmu
\author{Y.~Hayato}
\AFFicrr
\AFFipmu
\author{M.~Ikeda}
\AFFicrr
\author{K.~Iyogi}
\AFFicrr 
\author{J.~Kameda}
\author{Y.~Kishimoto}
\AFFicrr
\AFFipmu 
\author{Ll.~Marti}
\AFFicrr
\author{M.~Miura} 
\author{S.~Moriyama} 
\author{M.~Nakahata}
\AFFicrr
\AFFipmu 
% ...

\collaboration{The Super-Kamiokande Collaboration}
\noaffiliation